\documentclass[fleqn,usenatbib]{mnras}

\usepackage{newtxtext,newtxmath}
\usepackage{bm}

\usepackage[T1]{fontenc}
\usepackage{ae,aecompl}

\usepackage{graphicx}	
\usepackage{amsmath}	
\usepackage{amssymb}	
\usepackage[normalem]{ulem}

\newcommand{\LSST}{\textsc{lsst}}

\newcommand{\hMpc}{\ensuremath{~h^{-1}\mathrm{Mpc}}}

\newcommand{\degSq}{\ensuremath{~\mathrm{deg}^2}}
\newcommand{\map}[1]{\ensuremath{#1 \times #1 \degSq}}

\defcitealias{Takahashi2017}{T17}

\usepackage[dvipsnames]{xcolor}

\title[Weak lensing void finders]{Optimal void finders in weak lensing maps}

\author[C. T. Davies et. al]{Christopher T. Davies,$^{1}$\thanks{E-mail: christopher.t.davies@durham.ac.uk (CTD)}
Enrique Paillas,$^{2,3}$
Marius Cautun,$^{4}$
and Baojiu Li$^{1}$
\\
$^{1}$Institute for Computational Cosmology, Department of Physics, Durham University, South Road, Durham DH1 3LE, UK\\
$^{2}$Instituto de Astrof\'isica, Pontificia Universidad Cat\'olica de Chile, Av. Vicu\~na Mackenna 4860, Santiago, Chile\\
$^{3}$Centro de Astro-Ingenier\'ia, Pontificia Universidad Cat\'olica de Chile, Av. Vicu\~na Mackenna 4860, Santiago, Chile\\
$^{4}$Leiden Observatory, Leiden University, PO Box 9513, NL-2300 RA Leiden, the Netherlands \\
}

\pubyear{2020}

\begin{document}
\label{firstpage}
\pagerange{\pageref{firstpage}--\pageref{lastpage}}
\maketitle

\begin{abstract}
Cosmic voids are a key component of the large-scale structure that contain a plethora of cosmological information. Typically, voids are identified from the underlying galaxy distribution, which is a biased tracer of the total matter field. Previous works have shown that 2D voids identified in weak lensing maps -- weak lensing voids -- correspond better to true underdense regions along the line of sight. In this work, we study how the properties of weak lensing voids depend on the choice of void finder, by adapting several popular void finders. We present and discuss the differences between identifying voids directly in the convergence maps, and in the distribution of weak lensing peaks. Particular effort has been made to test how these results are affected by galaxy shape noise, which is a dominant source of noise in weak lensing observations. By studying the signal-to-noise ratios (SNR) for the tangential shear profile of each void finder, we find that voids identified directly in the convergence maps have the highest SNR but are also the ones most affected by galaxy shape noise. Troughs are least affected by noise, but also have the lowest SNR. The tunnel algorithm, which identifies voids in the distribution of weak lensing peaks, represents a good compromise between finding a large tangential shear SNR and mitigating the effect of galaxy shape noise.

\end{abstract}

\begin{keywords}
gravitational lensing: weak -- large-scale structure of Universe -- cosmology: theory -- methods: data analysis
\end{keywords}



\section{Introduction}
Gravitational lensing is the physical phenomena in which light is deflected by gravitational potentials along the line of sight, which results in the distortion and magnification of distant galaxy images. This phenomena can be split into two regimes, strong and weak gravitational lensing. For strong gravitational lensing, observed galaxy images are visibly distorted and multiple images of the same source galaxy can be produced. In the case of weak gravitational lensing (WL), where image distortions are very small, the underlying lensing signal can be recovered by statistically correlating distortions in many source galaxy images over extended patches of the sky \citep{Bacon2000,Kaiser2000,VanWaerbeke2000,Wittman2000}. In particular, WL is sensitive to moderate variations in the mass distribution, such as the large-scale structure (LSS) of the Universe, and allows us to map the cosmic mass content over a large range of scales, from kiloparsecs to hundreds of Megaparsecs \citep[see][for a review]{Bartelmann2001,Kilbinger2015}.

WL represents a powerful cosmological probe because it is an unbiased tracer of the cosmic LSS, whose properties and evolution are governed by the underlying cosmological model, including the matter content in the Universe and the law of gravity. Thus, WL can be used to constrain cosmological parameters within the standard $\Lambda\rm{CDM}$ paradigm, as well as models beyond $\Lambda\rm{CDM}$ \citep{Albrecht2006,LSST2012,Amendola2013,Weinberg2013}. In order to achieve this, one must construct statistics which efficiently capture the cosmological information embedded within WL maps. This can be achieved through two-point statistics such as the power spectrum or the two-point correlation function. One such example is the shear-shear correlation function which has been used to provide constraints on cosmological parameters within $\Lambda$CDM \citep[e.g.][]{Schneider2002,Semboloni2006,Hoekstra2006,Fu2008,Heymans2012,Kilbinger2013,Hildebrandt2017}. The convergence power spectrum and shear-shear correlation have also been used to test modified gravity theories beyond $\Lambda\rm{CDM}$ \citep[e.g.][]{Schmidt2008,Tsujikawa2008,Huterer2010}. 

The power spectrum encapsulates all the information required to describe a Gaussian random field, which is an accurate representation of the matter distribution in the Universe at early times. However, the growth of LSS is governed by gravity which induces non-Gaussian features due to nonlinear evolution at late times, when the power spectrum becomes an incomplete description of the underlying matter field. Therefore, for non-Gaussian observables such as WL maps, it is important to develop complementary statistics beyond the power spectrum in order to maximise the cosmological information that can be extracted. 

A popular and simple alternative WL statistic that is complementary to the WL power spectrum is the abundance of WL peaks \citep{Jain2000,Pen2003,Dietrich2010}, which are usually defined as the local maxima in the convergence field. The strongest WL peaks are typically produced by the most massive structures in the universe, such as galaxy clusters \citep{Yang2011,X.Liu2015,J.Liu2016}, and so the abundance of these WL peaks is directly sensitive to the non-Gaussian features of the cosmic web. Furthermore, low amplitude WL peaks have been shown to contain useful cosmological information \citep{Dietrich2010, Kratochvil2010, Yang2011}, making the study of weak lensing peaks crucial for cosmological constraints. This complementary information contained in the abundance of WL peaks has been exploited to improve cosmological constraints on $\Lambda$CDM parameters \citep{Shan2012,VanWaerbeke2013,Shan2014,X.Liu2015}, modified gravity \citep{Cardone2013,X.Liu2016,Higuchi2016,Shirasaki:2016twn,Peel2018}, dark energy \citep{Giocoli2018}, and the sum of neutrino masses \citep{Li2018}. Additional WL peak statistics, such as the two point correlation function, have also been shown to be sensitive to the $\Lambda\rm{CDM}$ parameters \citep{Davies2019}.

There are multiple other WL statistics beyond the power spectrum that have been utilised to constrain cosmology, and we briefly mention a few here. The first is Minkowski functionals, which can provide additional constrains on the dark energy equation of state parameter \citep{Kratochvil2012,Petri2013,Ling2015,Marques2018}. The WL bispectrum, which is sensitive to non-Gaussianity by definition, has been shown to be a useful statistic for future surveys \citep{Cooray2001,Rizzato2019,Munshi2019}, and can be used to improve parameter constraints, such as neutrino masses \citep{Coulton2018}. And finally, WL minima, local minima in the convergence field, are less sensitive to baryonic effects, and offer certain advantages over WL peaks \citep{Coulton2019}. Every such novel statistic offers its own unique advantages, which makes the study of novel statistics crucial.

The goal of this paper is to explore the properties of another of such statistic, WL voids, first introduced
in \cite{Davies2018}. Typically voids are identified in the full 3D distribution of the LSS, as regions with low densities of matter or tracers. The void abundance, their radial profiles and shapes contain higher order clustering information (and hence non-Gaussian information; \citealt{White1979,Fry1986,Biswas2010,Bos2012,Lavaux2012}). 
Most studies have focused on galaxy voids, which corresponds to underdensities in the galaxy distribution \citep[e.g.][]{Paz2013,Sutter2014,Cautun2016,Nadathur2016}. The statistics of galaxy voids contain complementary information to the galaxy power spectrum and baryonic acoustic oscillations \citep[e.g.][]{Pisani2015,Hamaus2016,Nadathur2019}. One useful void statistic is their WL profiles, which have been argued to represent a powerful cosmological probe \citep{Cai2015,Barreira2015,Falck2018}. 

Compared with galaxy voids, WL voids have been shown to corresponds to deeper line-of-sight projected underdensities and thus they have a larger tangential shear signal \citep{Davies2018}. This potentially makes WL voids better cosmological probes than galaxy voids. This has been exemplified by \citet{Davies2019b} in the context of a class of modified gravity models, which can be considerably better constrained with 2D WL voids than with galaxy voids. 

The total SNR of void lensing profiles depends on the number of voids and the amplitude of the lensing profile. Depending on how voids are identified, either fewer or more 2D voids can be obtained relative to 3D voids. However, most importantly, the 2D void lensing profiles have amplitudes roughly an order of magnitude larger than those of 3D voids \citep{Cautun2018,Davies2018}. This is the most important factor that contributes to higher SNR for 2D WL voids compared to 3D voids in the cosmic web.

\citet{Davies2018} focused on a particular class of WL voids, called VOLEs (VOids from LEnsing), where the voids are identified as circles devoid of weak lening peaks. However, as for 3D voids, the definition and therefore the finding algorithm of 2D voids are not unique. There are multiple methods of finding underdensities, and thus multiple approaches to define voids \citep[e.g.][]{Colberg2008,Cautun2018}. This ambiguity can lead to systematic differences in void observables among the various void finders. However this ambiguity can also be exploited, by picking the void-finding algorithm that best suits the intended purpose. In our case, we want to maximise the amplitude of the WL void lensing profiles (or similarly the SNR of the WL void lensing profiles), whilst also limiting the impact of observational noises on the resulting void statistics. To this end, we will present WL void statistics for a range of void-finding algorithms, and discuss the limitations and advantages of each void finder.

Here, we compare seven different void definitions. These can be split into two classes. First and seemingly the most natural approach, consists of the methods which identify voids directly from the WL convergence field. In the following, we denote the convergence with $\kappa$. The simplest objects that can be considered as WL voids are the WL minima (i.e., local minima in the $\kappa$ field) where the deepest minima have been shown to correspond to large supervoids along the line of sight \citep{Chang2018}. More advanced void definitions include the watershed void finder (WVF; \citealt{Platen2007}), which identifies voids as the watershed basins of the convergence field, the spherical void finder (SVF; e.g., \citealt{Padilla2005}) applied to the convergence field (which we denote as SVF $\kappa$), which finds the largest circles whose mean $\kappa$ is below a given threshold, and troughs (denoted with Troughs $\kappa$; \citealt{Gruen2015}), which consists of fixed sized circles whose mean convergence is below a given threshold. 

By construction, the number and properties of voids identified in the convergence field are sensitive to the lowest $\kappa$ values. These regions are the ones affected the most by galaxy shape noise (GSN). For this reason we consider a second class of void finders, which consists of methods that identify voids using a distribution of tracers, which we take to be the peaks of the convergence field (as we shall discuss, the peaks are less affected by GSN). We study three methods in this class: the `tunnel' algorithm \citep{Cautun2018} employed in \citet{Davies2018}, which identifies voids as the largest circles devoid of tracers, the SVF but now applied to the peak distribution (hereafter referred to as `SVF peak'), and troughs identified in the peak distribution (denoted with `Troughs peak'), which consists of fixed sized circles that enclose fewer than a given number of peaks. A detailed description of how each WL void finder is presented in Section \ref{sec:void finders}.

The content of the paper is as follows: in Section \ref{sec:Theory} we present the relevant WL theory. The numerical simulations and galaxy shape noise prescription used in this study are presented in Section \ref{sec:Weak lensing maps} along with the basic WL map statistics which will help the interpretation of results from different WL void finders. The void finders studied here are presented in Section \ref{sec:void finders}, and the statistics describing the WL voids associated to each WL void finder are presented and discussed in Section \ref{sec:void statistics}. We then compare useful properties of the WL void finders in Section \ref{sec:comparison}, with the discussion and conclusions in Section \ref{sec:discussion and conclusions}. We also present the correlation matrices of the tangential shear profiles for different void finders in Appendix \ref{app:correlation}. In Appendix \ref{app:WL voids in GSN maps} we test how WL voids behave in WL maps with only GSN i.e. WL maps with no physical signal, and discuss how WL voids are sensitive to the physical information in WL maps.

\section{Theory}\label{sec:Theory}

For a gravitationally lensed image, the lens equation is given by
\begin{equation}
    \pmb{\alpha} = \pmb{\beta} - \pmb{\theta} \, ,
\end{equation}
where $\pmb{\theta}$ is the observed position of the lensed image, $\pmb{\beta}$ is the true position of the source on the sky, and $\pmb{\alpha}$ is the deflection angle.
The deformation matrix \textbf{A} can be defined as
\begin{equation}
    A_{ij} = \frac{\partial \beta_{i}}{\partial \theta_{j}} = \delta_{ij} - \frac{\partial \alpha_{i}}{\partial \theta_{j}} \, ,
    \label{eq:amp mat}
\end{equation}
while, under the Born approximation and neglecting lens-lens coupling, the deflection angle can be expressed as the gradient of a 2D lensing potential, $\psi$, which is given by
\begin{equation}
    \psi(\pmb{\theta},\chi) = \frac{2}{c^2} \int_0^{\chi} \frac{\chi - \chi'}{\chi \chi'} \Phi(\chi' \pmb{\theta},\pmb{\theta}) d\chi' \, .
    \label{eq:lensing potential}
\end{equation}
Here, $\chi$ is the comoving distance to the source, $\chi'$ is the comoving distance to the lens, $c$ is the speed of light and $\Phi$ is the 3D lensing potential of the lens. In the absence of the anisotropic stress, which means that the two gravitational potentials in the Newtonian gauge are both equal to $\Phi$, $\Phi$ is related to the non-relativistic matter density contrast, $\delta$, through the Poisson equation 
\begin{equation}
    \nabla^2\Phi = 4 \pi G a^2 \bar{\rho} \delta \, ,
    \label{eq:Poisson equation}
\end{equation}
where $G$ is the gravitational constant, $a$ is the scale factor, $\bar{\rho}$ is the mean matter density of the universe, and $\delta = \rho/\bar{\rho} - 1$. Eq.~\eqref{eq:lensing potential} shows that the WL signal is produced by the matter distribution along the entire line of sight from the source to the observer.

Using $\pmb{\alpha} = \pmb{\nabla}\psi$ allows Eq. \eqref{eq:amp mat} to be expressed in terms of $\psi$
\begin{equation}
    A_{ij} = \delta_{ij} - \partial_{i} \partial_{j} \psi \, ,
\end{equation}
where partial derivatives are taken with respect to $\pmb{\theta}$. The $\pmb{A}$ matrix can be parameterised in terms of convergence, $\kappa$, and shear, $\gamma = \gamma_1 + i\gamma_2$, as
\begin{equation}
    \pmb{A} = 
    \begin{pmatrix}
    1 - \kappa -\gamma_1 & -\gamma_2\\
    -\gamma_2 & 1-\kappa+\gamma_1
    \end{pmatrix}
    \, ,
\end{equation}    
where the convergence and shear are related to the lensing potential via
\begin{equation}
    \kappa \equiv \frac{1}{2} \nabla^2_{\pmb{\theta}} \psi \, ,
    \label{eq:convergence}
\end{equation}
\begin{equation}
    \gamma_1 \equiv \frac{1}{2}\left(\nabla_{\pmb{\theta}_1}\nabla_{\pmb{\theta}_1}-\nabla_{\pmb{\theta}_2}\nabla_{\pmb{\theta}_2}\right)\psi, 
    \quad\quad\quad
    \gamma_2 \equiv \nabla_{\pmb{\theta}_1}\nabla_{\pmb{\theta}_2}\psi,
    \label{eq:shear}
\end{equation}
where $\nabla_{\pmb{\theta}} \equiv (\chi')^{-1}\nabla$. Eq.~\eqref{eq:convergence} can be interpreted as a 2D Poisson equation, and so by substituting Eq.~\eqref{eq:Poisson equation} and Eq.~\eqref{eq:lensing potential} into Eq.~\eqref{eq:convergence}, the convergence can be expressed in terms of the matter overdensity
\begin{equation}
    \kappa(\pmb{\theta},\chi) = \frac{3H_0^2\Omega_{\rm{m}}}{2c^2}\int_0^{\chi}\frac{\chi - \chi'}{\chi} \chi' \frac{\delta(\chi'\pmb{\theta},\chi')}{a(\chi')} d\chi' \, .
    \label{eq:conv source}
\end{equation}
This shows that the observed WL convergence can be interpreted as the projected density along the line of sight, weighted by the lensing efficiency factor $(\chi-\chi')\chi'/\chi$.

In WL observations, the source galaxies do not occupy a single plane at a fixed distance from the observer. The observed catalogue of source galaxies has a probability distribution $n(\chi)$, and Eq. \eqref{eq:conv source} must be weighted by this source galaxy distribution in order to obtain $\kappa(\pmb{\theta})$ \citep[see, e.g.,][for a more detailed discussion.]{Kilbinger2015}
\begin{equation}
    \kappa(\pmb{\theta}) = \int_0^{\chi} n(\chi') \kappa(\pmb{\theta},\chi') d\chi' \, .
\end{equation}

Finally, we can relate the radial convergence profile of an object $\kappa(r)$ to its radial tangential shear profile through
\begin{equation}
    \gamma_{\rm{t}}(r) = \bar{\kappa}(< r) - \kappa (r)
    \label{eq:gamma_t} \;,
\end{equation}
where
\begin{equation}
    \bar{\kappa}(< r) = \frac{1}{\pi r^2}\int_0^{r} 2 \pi r' \kappa(r') dr'
    \;
\end{equation}
is the mean enclosed convergence within radius $r$. Notice that here and throughout this paper we use $r$ rather than $\theta$ to represent the 2D distance from the void centre.

In addition to the convergence profiles of WL voids, it is useful to also study the tangential shear profiles, since the tangential shear is the quantity directly measured by observations.

\section{Weak lensing maps}
\label{sec:Weak lensing maps}

In this section, we briefly outline the numerical simulations and the weak lensing maps used in this study, our prescription for including galaxy shape noise in our analysis, and a discussion on the relevant WL statistics that will inform the interpretation of our results from different void finders.

\subsection{Numerical simulations}

To study WL voids we use WL maps generated from N-body simulations taken from \cite{Takahashi2017} (herein \citetalias{Takahashi2017}) which provide publicly-available all-sky WL convergence maps. The WL maps are generated with the ray tracing algorithm from \citet{Hamana2015} \citep[see also][]{Shirasaki2015}. These WL convergence maps have a HEALPix resolution of $N_{\rm{side}}=16384$, and a source redshift of $z_{\rm{s}}=1$. The N-body simulations have a particle number of $2048^3$, and the particle mass varies with the box size ranging from $8.2\times10^8$ to $2.3\times10^{12}M_\odot$ (see Table 1 of \citetalias{Takahashi2017} for more details). To avoid repeating structures along the line-of-sight, \citetalias{Takahashi2017} constructed the light cone by stacking cubic simulation boxes of increasing size, with comoving sizes $L,2L,3L,\cdots,14L$, where $L=450h^{-1}$Mpc. These boxes are then duplicated 8 times and nested around the observer, where nests of larger boxes contain nests of smaller boxes at their centres. The matter distribution of these nested boxes is projected onto the nearest spherical shell centered on the observer, where the shells have radii of $ N \times 150 \hMpc$ with $N=1,\cdots,14$ (see \citetalias{Takahashi2017} for illustration). The cosmological parameters used for these WL maps corresponds to a flat universe with $\Omega_{\rm{m}} = 0.279$, $\Omega_\Lambda=0.721$, $\sigma_8 = 0.820$ and $h = 0.7$, where $h = H_0 / 100$ km s$^{-1}$ Mpc$^{-1}$.

We split the all sky WL convergence maps into 192 \map{10} maps and then extend the map boundaries by a further 5 deg on all sides giving us 192 \map{20} maps with a resolution of $4096^2$ pixels. This approach results in maps where the central \map{10} region of each map does not overlap with the central \map{10} region of any of the remaining 191 maps. The use of the 192 smaller maps allows us to stick to the flat sky approximation. Void detection is carried out on the full \map{20} and voids with centres outside of the central \map{10} are discarded. Additionally, voids that are within twice their radius from the map boundary are discarded when calculating the void lensing profiles. This approach guarantees that void identification is not biased away from large voids due to boundary effects. For more details on our projection method, see Appendix A of \cite{Davies2019}.

\subsection{Galaxy shape noise}
\label{sec:GSN prescription}

The observed correlation in galaxy shapes induced by gravitational lensing is entirely dominated by the random shapes and orientations of galaxies, which are referred to as galaxy shape noise (GSN). As shown by \cite{VanWaerbeke2000b}, GSN can be modelled by adding random values drawn from a Gaussian distribution to each pixel of our simulated WL maps. The standard deviation of this distribution is given by
\begin{equation}
    \sigma_{\rm{pix}}^2 = \frac{\sigma_{\rm{int}}^2}{2 \theta_{\rm{pix}} n_{\rm{gal}} }
    \;,
    \label{eq: GSN gaussian}
\end{equation}
where $\sigma_{\rm{int}}$ is the intrinsic ellipticity dispersion of the source galaxies, $\theta_{\rm{pix}}$ is the width of each pixel, and $n_{\rm{gal}}$ is the measured source galaxy number density. We use $\sigma_{\rm{int}} = 0.4$ and $n_{\rm{gal}} = 40 $ arcmin$^{-2}$, which match $\LSST$ specifications \citep{LSST2009}.

The inclusion of GSN results in noise-dominated WL maps. Nevertheless, the noise effect can be suppressed by smoothing with a (usually) Gaussian filter with smoothing length $\theta_{\rm{s}}$. Using a small value for $\theta_{\rm{s}}$ allows a given WL statistic to probe the smallest scales and maximise the information gained, however this also leaves significant contamination from GSN. Using larger $\theta_{\rm{s}}$ values reduces the GSN contamination, but suppresses the small scale information within the WL maps. This means that a trade off must be struck between sufficiently suppressing GSN and retaining WL information on small scales. Additionally, the analysis carried out here relies on WL maps generated from dark matter only simulations, and do not include baryon physics. To suppress the differences between dark matter only and full hydrodynamic simulations, \cite{Weiss2019} found that very large smoothing scales must be used. Furthermore, \cite{J.Liu2015} found that constraints on cosmological parameters from WL peaks are improved when multiple smoothing scales are used. These imply that there is no single best choice of smoothing scale that fits all purposes when analysing WL statistics. So in order to explore this fully, all statistics in this work will be shown for multiple smoothing scales, $\theta_{\rm{s}} = 1$ (blue), $2.5$ (orange), and $5$ (green) arcmin, both in the presence (dashed) and absence (solid) of GSN.

By presenting all statistics for multiple smoothing scales, with and without GSN, we will be able to identify the void finders that are the least affected by GSN. However at this point the impact of GSN on cosmological parameter constraints from WL voids is not known. It is possible that the inclusion of GSN may improve cosmological parameter constraints from WL voids by increasing the signal-to-noise (SNR) ratio relative to the case where GSN is not included, as has been found with WL peaks \citep{Yang2011}. However, GSN could also bias or degrade the cosmological parameter constraints from WL voids. We leave such an investigation to further work and focus on identifying void finders that are the least affected by GSN in this paper. 

For the analysis of WL peaks it is useful to define the amplitude of a given peak relative to the r.m.s.~fluctuation of the added GSN component of the WL field. As such $\nu$ is defined as
\begin{equation}
    \nu \equiv \frac{\kappa}{\sigma_{\rm{GSN}}(\theta_{\rm{s})}} \, ,
\end{equation}
where $\sigma_{\rm{GSN}}(\theta_{\rm{s}})$ is the standard deviation of the smoothed GSN map (without contributions from the physical WL convergence map i.e. noise only) and varies depending on the smoothing scale with which the WL peak is identified, with $\sigma_{\rm{GSN}} = 0.0126,0.0051$ and $0.0025$ for $\theta_{\rm{s}} = 1, 2.5$ and $5$ arcmin respectively.

\subsection{Convergence PDF and WL peak abundance}

\begin{figure*}
    \centering
    \includegraphics[width=2\columnwidth]{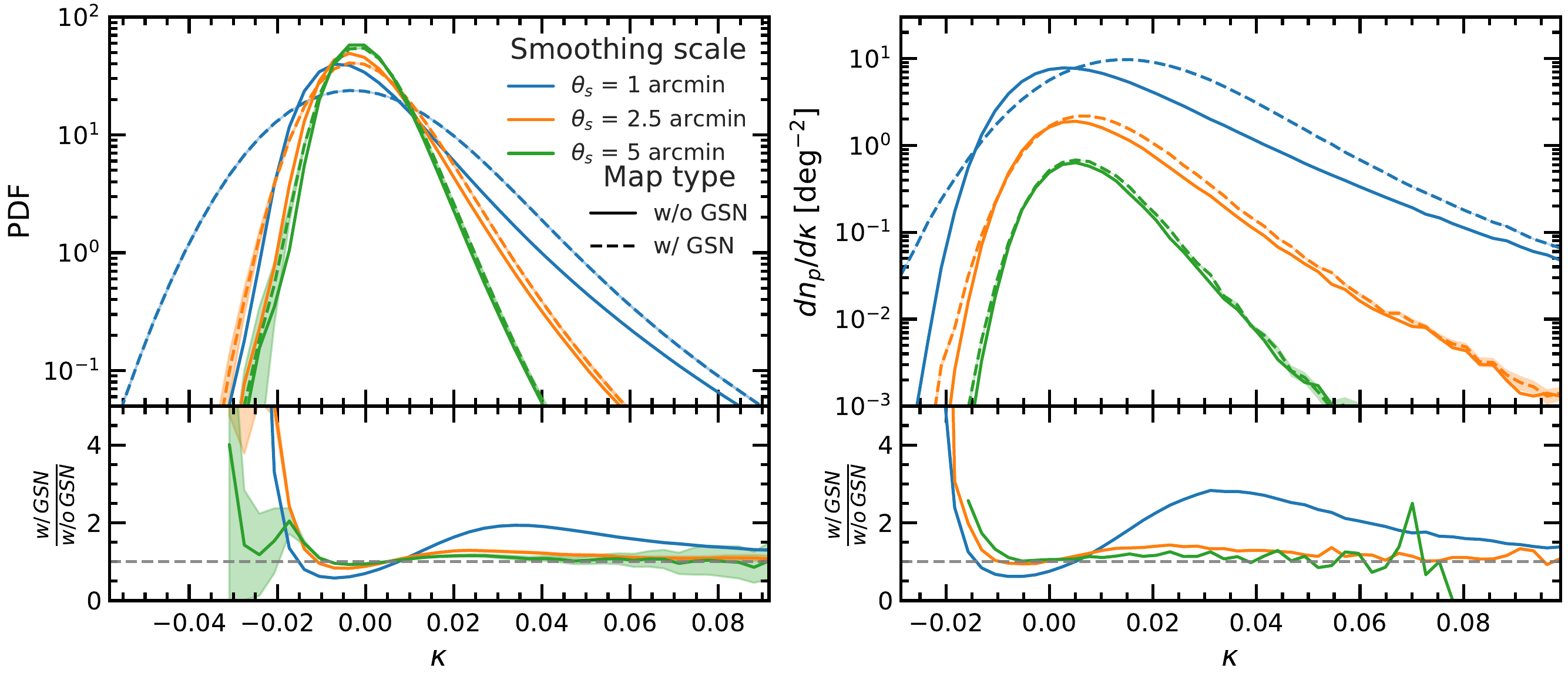}
    \caption{Left panels: the probability distribution function (PDF) of the WL convergence field, $\kappa$. Right panels: The differential abundance of WL peaks as a function of peak height $\nu$. The results  shown here are obtained using a ${\sim}19,000\degSq$ area with the shaded regions denoting the one sigma error bars (most of the time the errors are smaller than the line thickness). The dashed and solid lines correspond to the WL convergence maps with and without GSN respectively. The colours correspond to different smoothing scales of the $\kappa$ field: $1.0$ (blue), $2.5$ (orange) and $5.0$ (green) arcmin. The relative differences between the cases with and without GSN are shown in the lower sub-panels.
    \newline
    }
    \label{fig:kappa pdf and WL peak abundance}
\end{figure*}

In order to aid the interpretation of the various WL void statistics, we first present some simple statistics that describe the information given to the WL void finders. In the cases of void finders applied directly to the convergence field this is the WL convergence probability distribution function (PDF) shown in the left panel of Fig.~\ref{fig:kappa pdf and WL peak abundance}, and for the void finders that use weak lensing peaks as tracers this is the WL peak abundance shown in the right panel Fig.~\ref{fig:kappa pdf and WL peak abundance}. Note that we define a WL peak as a pixel with a convergence value larger than that of its eight neighbours.

The left panel of Fig.~\ref{fig:kappa pdf and WL peak abundance} shows the WL convergence PDF for the three smoothing scales (1, 2.5 and 5 arcmin), for cases with and without the inclusion of GSN (dashed and solid). The convergence PDF is well described by a log normal distribution convolved with a Gaussian when GSN is included \citep{Clerkin2017}. The different colours show that as the smoothing scale increases, the width of the distribution decreases, suppressing the non-Gaussian structures within the WL map, and the agreement between the cases with and without GSN improves. The relative differences in the convergence PDF between the no-GSN and the GSN-added cases are larger for $\kappa<0$ than for $\kappa>0$, as can be seen more clearly in the lower panel. Therefore in order to find agreement in WL void statistics with and without the inclusion of GSN we will likely require larger smoothing scales than what is required to get the same agreement for WL peak statistics. Finally, for a smoothing scale of 1 arcmin (blue curves), the inclusion of GSN introduces a significant number of negative convergence values that are much lower than the lowest convergence values found in the WL maps without GSN. This indicates that 1 arcmin smoothing might be too small for WL void finders applied directly to the convergence field in order to agree before and after GSN is added. However, agreement between the two cases is largely improved once the smoothing scale is increased to 2.5 or 5 arcmin.

The differential WL peak abundances identified from WL maps with and without GSN smoothed over the three smoothing scales (1, 2.5 and 5 arcmin) are displayed in the right panel of Fig.~\ref{fig:kappa pdf and WL peak abundance}. By adding GSN, the peak of the distribution is shifted to the right, and more peaks are created. The addition of these spurious peaks from GSN will lead to the identification of spurious voids for void finders that find voids in the WL peak distribution. The differences between WL peak catalogues for maps with and without GSN is suppressed as the smoothing scale increases, but this also decreases the overall abundance of the WL peaks. It can also be seen that, as $\kappa$ increases, the differences between the maps with and without GSN decreases. This is because the largest WL peaks are less affected by GSN, since the physical peak signal dominates over the noise.

The right panel of Fig.~\ref{fig:kappa pdf and WL peak abundance} also shows that there are many WL peaks with negative convergence values, which are local maxima in underdense regions of the WL convergence maps. This is as expected, since most regions have $\kappa<0$ (see left panel in Fig.~\ref{fig:kappa pdf and WL peak abundance}) and thus many local maxima will have heights $\kappa<0$. As we will discuss in Section~\ref{sec:void finders}, the void finders based on the peak distribution identify the voids as the regions that are largely devoid of peaks. Including all the WL peaks in our analysis can raise two problems. Firstly, it reduces the contrast in peak number density between overdense and underdense regions, and thus makes it difficult to robustly identify the underdense regions. Secondly, the location and height of $\kappa\lesssim0$ peaks is much more affected by GSN than for the high $\kappa$ peaks. This defeats the main reason for identifying voids using the WL peaks, which is to mitigate the effect of GSN on the WL void population. Therefore, to deal with these two issues, we proceed by imposing a peak height cut on the WL peak catalogues, and remove all peaks below a given threshold. This adds a free parameter to the analysis and thus, for the void finders that use WL peaks as tracers, we will present results for peak catalogues with peak heights of $\nu > 2$ and $\nu > 4$.

\section{Void finders} 
\label{sec:void finders}

In this section, we describe the implementation of each WL void finder used in this paper. These void finders were originally developed to identify voids in a 3D galaxy or matter distribution, which means that some must be modified slightly to identify 2D WL voids. In each case we try to minimise the extent of the modification so that the interpretation of results can remain as similar as possible to the interpretation of 3D voids. Furthermore, where possible, we apply each void finder to both the WL peak distribution and the WL convergence field to see which approach provides the most information (in terms of the signal-to-noise ratio, SNR) and which is least affected by GSN. Finally, all void identification is carried out on the full \map{20} maps, while the voids whose centres reside outside of the central \map{10} are discarded. This ensures that the void identification process is not contaminated by edge effects, and that we do not bias our results away from large voids, since larger voids are more likely to intersect the map boundary. 

\subsection{Minima}

Weak lensing minima are the simplest objects which can be interpreted as WL voids, which correspond to the most underdense lines of sight within the WL convergence maps. Here we define WL minima as local minima in the convergence field, which is a pixel whose $\kappa$ value is lower than those of its eight neighbours. We identify WL minima in the smoothed convergence field and discard all minima with a positive $\kappa$ value, because a positive $\kappa$ value indicates that the minimum and its neighbours reside within a local overdensity. This allows us to remain consistent with the general definition of a WL void, which is an underdense patch of the WL convergence map.

\subsection{Troughs}

Troughs \citep{Gruen2015} are underdense circles of a fixed size. Typically troughs are identified by randomly placing circles of that fixed size in a projected galaxy field and discarding the circles that contain the most galaxies, leaving only those that contain the least galaxies. Here we adapt the trough algorithm and apply it to both the WL peak field and the WL convergence field.

For troughs applied directly to the convergence field (Troughs $\kappa$), we first place $5000$ circles\footnote{We have also run the trough algorithm with 10 times as many randomly placed circles, and find that this does not change the SNR values of the trough tangential shear profiles. Therefore in this paper we stick to 5000.} randomly such that their centres fall into the central \map{10} of the WL convergence map. For each of these circles, the mean enclosed convergence is calculated. The trough catalogue consists of the $20\%$ of the circles with the lowest mean enclosed convergence. The above procedure is carried out for circles with radii of $10$, $20$ and $30$ arcmin, which correspond to the typical values used in previous studies \citep[e.g.,][]{Barreira2017,Gruen2018}. 

For troughs identified in the WL peak distribution (Troughs peak), the same steps are repeated except that, rather than calculating the mean enclosed convergence, we count the number of enclosed peaks, and keep the $20\%$ of circles which contain the fewest peaks. Again, these steps are repeated for circles of radii of $10$, $20$ and $30$ arcmin. 

The number of randomly placed circles and the upper fraction of circles to be discarded are both free parameters. However, to keep the analysis in this work simple we do not vary these parameters, and their values above have been chosen to match the typical abundances of WL voids produced by the other algorithms for a fair comparison. 

\subsection{Watershed void finder (WVF)}

The watershed void finder \citep[][WVF]{Platen2007} defines voids as the watershed basins, which are analogous to water basins formed from rain running down a landscape. To identify the watershed basins, each pixel of the convergence map is connected to its neighbour with the lowest density, and this process is repeated for successive neighbours until a local minima is reached. All pixels connected to the same minima then belong to the same watershed basin. This results in ridges of local overdensities along the basin boundaries. 

To mitigate the impact of GSN, we compare the average amplitude of each basin boundary with the amplitude of their corresponding minima. If the absolute difference in amplitude between the two is less than $h_{\rm{boundary}}$, we merge that basin with its neighbour, which creates a single larger basin. In this analysis we choose $h_{\rm{boundary}} = \sigma_{\rm{GSN}} / 2$, which allows watershed basins that have been artificially split by spurious structures introduced by GSN to be re-merged. Adding the basin merge criteria means that $h_{\rm{boundary}}$ is an additional free parameter in the WVF algorithm. We have tested the impact of varying $h_{\rm{boundary}}$ and find that it has little impact on our results. We choose $h_{\rm{boundary}} = \sigma_{\rm{GSN}} / 2$ as a compromise between mitigating the impacts of GSN on watershed basins and over merging, which would on average flatten out void lensing profiles. 

This algorithm generates irregular basins which span the entire area of the WL convergence map. In order to calculate the stacked lensing profiles of the voids, we must define their void centres and radii using the information of the corresponding basins. We take the void centres as the area-weighted barycentre of all the pixels in each basin and define an effective void radius of $R_{\rm{v}} = (A/\pi)^{1/2}$ (which is the radius of a circle with the same area $A$ as the irregular basin) when calculating the WVF lensing profiles. 

When the watershed algorithm is applied to the galaxy distribution to find 3D voids in the LSS, the galaxies are first used as tracers to construct an estimate of the underlying density field using a Delaunay tessellation field estimation (DTFE) \citep{Schaap2000,Cautun2011}. This in principle means that WL peaks could also be used to identify WL voids with the watershed algorithm, by using the WL peaks to construct a WL peak density field. However, we find that the usual DTFE approach is insufficient, since it results in WL voids identified from the WL peak distribution that bear little correlation to underdensities in the original convergence map. While it may be possible to improve this procedure by using information about the WL peak heights in the DTFE reconstruction, this is beyond the scope of this work, and we thus instead choose to only study voids identified by applying the watershed algorithm to the WL convergence field. 

\subsection{Spherical void finder (SVF)}

The spherical void finder (SVF) \citep[e.g.,][]{Padilla2005} identifies underdense spherical regions in the galaxy distribution, by growing spheres around regions that are empty of galaxies. When applied to find WL voids, the SVF identifies circular regions in the WL convergence or peak fields that are below a specified `density' threshold. In practice, in order to allow SVF voids to `grow' as large as possible, circles are shrunk from some arbitrarily large size around candidate void centres until the threshold is met. 

For the SVF applied directly to the WL convergence map (SVF $\kappa$), local minima are considered as prospective void centres. Starting from a large radius, circles are then shrunk around these void centres until the mean enclosed convergence reaches a predefined threshold, $\kappa_{\rm{thresh}}$. Here, larger values of $\kappa_{\rm{thresh}}$ result in larger voids, and note that we require $\kappa_{\rm{thresh}}$ to be negative so that the SVF finder identifies regions that enclose underdense sections of the convergence map. We have tested a range of values for $\kappa_{\rm{thresh}}$, and as a compromise between identifying the most underdense regions and allowing voids to grow as large as possible, we set $\kappa_{\rm{thresh}} = -0.01$ in this analysis. Once all prospective voids are shrunk until their mean convergence is $\kappa_{\rm{thresh}}$, we proceed to remove the objects that overlap significantly. That is, if the distance between any two prospective voids is less than half the sum of their radii, we discard the smaller of the two. Finally, we remove all voids with radii less than twice the smoothing scale that is applied to the convergence map ($\theta_{\rm{s}}$) to reduce the number of spurious voids.

For the SVF applied to the WL peak distribution (SVF peak), a Delaunay triangulation of the peak field is performed, and the circumcentres associated to each triangle are considered as potential void centres. Starting from a large radius, circles around those centres are shrunk, until the mean enclosed WL peak number density reaches a predefined fraction of the mean WL peak number density. We find that the resulting void catalogues are somewhat insensitive to the exact choice of this threshold value, and therefore pick $40\%$ as a good compromise between allowing SVF voids to grow as large as possible and ensuring these voids correspond to underdense regions of the WL convergence maps. Next, we randomly shift void centre positions within the void radius, in order to verify if it is possible for the void to `grow' a bit more (i.e., to reach the density threshold at a slightly larger radius). Finally, if the centres of two voids are separated by less than half of the sum of their radii, we remove the smaller of the two.

\subsection{Tunnels}

The tunnel algorithm \citep{Cautun2018} identifies the largest circles in a 2D tracer catalogue that are empty of tracers. Initially, a Delaunay tessellation with WL peaks as the vertices is constructed. This produces a tessellation of Delaunay triangles, with a WL peak at the corner of each triangle, and no WL peaks within the triangles. Each Delaunay triangle is then used to construct its corresponding circumcircle, which is the circle that resides directly on top of the Delaunay triangle, with the three vertices of the triangle falling on the circumcircle's circumference. This unique tessellation, by definition, produces circles which do not enclose any WL peaks. To avoid highly overlapping objects, we discard any tunnels whose centres reside within a larger tunnel. Recently, \citet{Davies2018} have studied tunnels in the context of WL maps and \citet{Davies2019b} have shown that they are better at constraining a modified gravity model than tunnels identified in the projected galaxy distribution.

\subsection{Visualisation}

\begin{figure*}
    \centering
    \includegraphics[width=2\columnwidth]{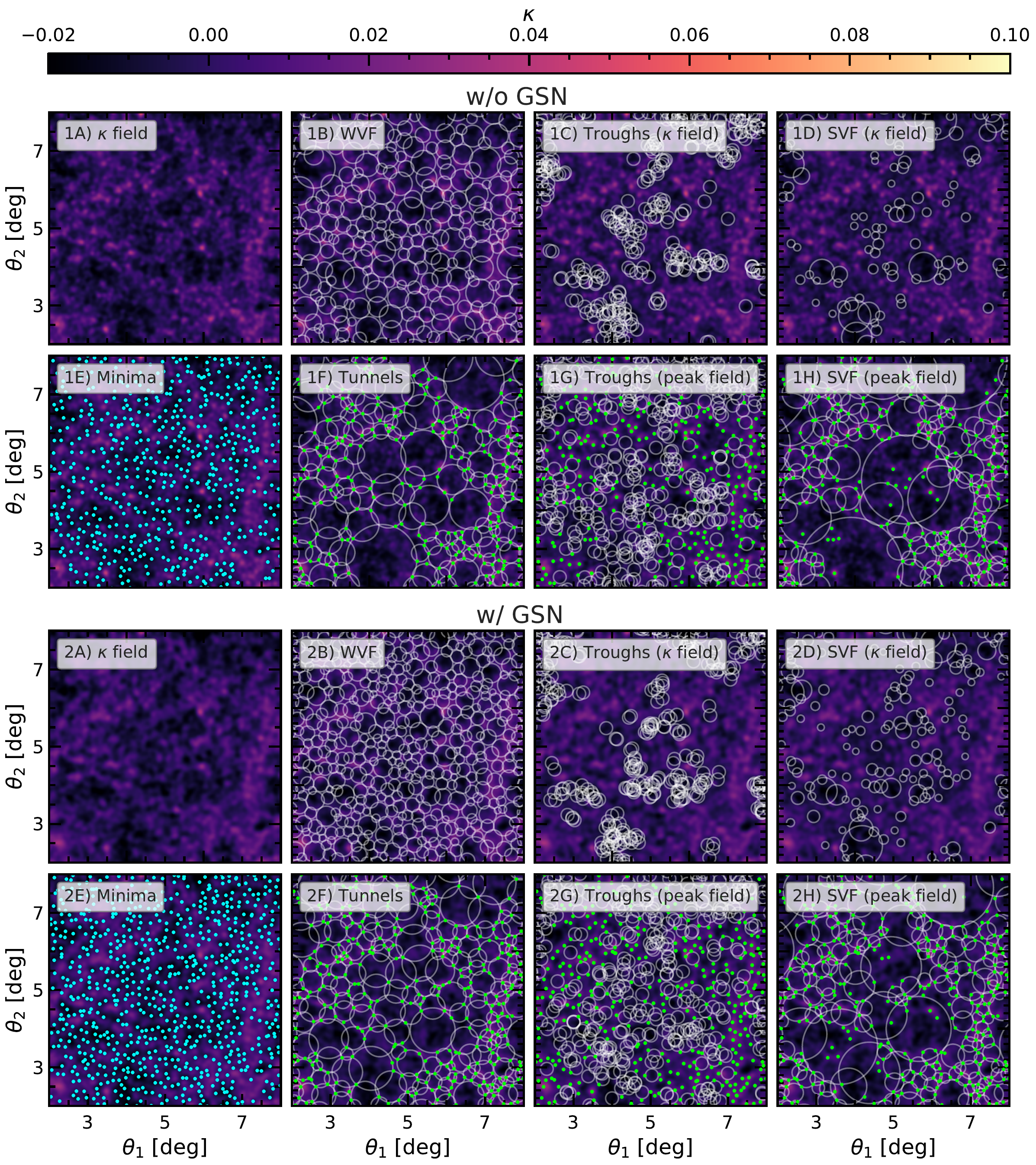}
    \caption{A visualisation of the weak lensing void finders discussed in this work. The convergence field is shown by the background colour map in each panel, with the convergence values illustrated by the colour-bar at the top of the figure. Here the brightest (orange) colours correspond to high $\kappa$ values and the darkest (purple) colours show low $\kappa$ regions. The results presented here are for a Gaussian smoothing scale, $\theta_{\rm{s}} = 2.5$ arcmin. The top eight panels are for WL maps with no GSN (1A to 1H), and the bottom eight panels are for WL maps with GSN (2A to 2H). Panels 1A and 2A show only the convergence fields as a reference point. The panels 1B to 1E and 2B to 2E show voids identified in the convergence field and correspond to: WVF, troughs and SVF applied to the $\kappa$ field, and minima. The remaining panels (1F to 1H and 2F to 2H) show voids identified using WL peaks with height, $\nu > 2$, and correspond to: tunnels, troughs and SVF applied to the peak distribution. Only the central \map{6} of the convergence maps are shown to avoid overcrowding. 
    }
    \label{fig:visualisation}
\end{figure*}

Fig.~\ref{fig:visualisation} shows a visualisation of each of the void finders studied in this work. The eight panels in the top section (1A -- 1H) show results for WL maps without GSN and the eight panels in the bottom section (2A -- 2H) are results for WL maps with GSN. Each panel corresponds to a different void finder, apart from the first panels of each section (panel 1A and 2A) which shows only the WL convergence field for reference. Only the central \map{6} of one of the maps are shown, to avoid over crowding whilst still displaying a fair sample of each void catalogue. The results shown here are for a smoothing scale of $\theta_{\rm{s}} = 2.5$ arcmin and for peak catalogues with WL peak heights of $\nu > 2$ (where applicable). The top row of each section (panels 1A - 1D and 2A - 2D) corresponds to voids identified in the WL convergence maps and the bottom rows (panels 1E - 1H and 2E - 2H) corresponds to voids identified in the WL peak distribution. The WL peaks are shown by the green points, while the WL minima are shown by the cyan points.

Panels 1B and 2B of Fig.~\ref{fig:visualisation} shows the WVF voids identified in the WL convergence map. These voids tend to avoid the more overdense patches of the convergence map, since these more overdense regions reside at the watershed basin boundaries. The WVF voids occupy most of the area of the WL convergence map, which is due to every pixel within the map being assigned to a watershed basin. In some cases, the largest voids enclose smaller voids, as can be seen towards the top left of Panel 1B. The overlap is an artefact of illustrating the WVF as circles when actually these voids have highly non-circular and non-overlapping shapes \citep[e.g. see][]{Platen2007,Cautun2016}. By adding GSN, the size of the WVF voids is reduced and their abundance is increased. 

Troughs identified directly on the convergence map are shown in Panels 1C and 2C, where it can be seen that these troughs trace only the most underdense regions of the convergence maps, which is by construction. The consequence of this algorithm is that many troughs significantly (or nearly entirely) overlap with other troughs, with very few troughs existing in isolation from other troughs. This will lead to highly correlated information in the statistics describing these troughs, as will be seen in their correlation matrices in Appendix \ref{app:correlation}. Panel 2C shows how adding GSN can change the spatial distribution of the troughs, although the degree of overlap between neighbouring troughs remains similar to the no-GSN case in panel 1C. 

Panels 1D and 2D show the SVF voids identified in the convergence field. As can be seen there, the abundance of these voids is significantly lower compared to void catalogues from other algorithms, and more small voids are generated. However, these voids trace the underdensities of the convergence map reasonably well, as can be seen by their dark interiors. There are more voids in panel 2D, indicating that GSN increases the abundance of these voids.  

The WL minima are displayed in Panels 1E and 2E. We remind the reader that we only study underdense minima, i.e., $\nu < 0$, and so only these minima are shown in the figure. These panels illustrate that the WL minima are slightly different from the typical WL void definition used in this work, since they have no size or radius, which has the advantage of simplicity. In later sections we'll discuss the abundance of WL minima as a function of their amplitude, rather than as a function of their size, and the abundance of WL minima has been shown to provide complementary cosmological information to the WL peak abundance \citep{Coulton2019}. We also discuss, for the first time, the potential for the radial lensing profiles of WL minima to be used in a cosmological analysis. There are more WL minima in panel 2E compared to 1E, indicating that there are more spurious minima created by GSN than physical minima removed by GSN. 

A visualisation of the tunnel algorithm is shown in Panels 1F and 2F. The WL peaks used to identify the tunnels are shown by the green points, highlighting that the tunnels do not enclose any WL peaks, and that the peaks only reside on the void boundaries. Like the WVF, the tunnels occupy most of the area of the convergence map, however the tunnel algorithm identifies a wider range of void sizes, producing more large voids than those identified in the convergence maps. Smaller tunnels tend to cluster more than the larger ones, with the former appearing more in the overdense parts of the convergence map. Also similar to the WVF voids, panel 2F contains more tunnels which are on average smaller than the tunnels in panel 1F. This is because the spurious WL peaks created by GSN break up the larger tunnels in panel 1F into the multiple smaller tunnels seen in panel 2F. 

Panels 1G and 2G show the troughs identified in the WL peak distribution. The troughs identified in this way still have a significant degree of overlap, however the overlap in this case is much weaker than for the troughs identified in the convergence maps. There are underdense patches in which no troughs have been placed, whilst many overlapping troughs can be seen in other regions. This highlights the inefficiency of the trough algorithm when applied to a WL peak distribution. This may be solved by increasing the number of troughs that are placed, however this will also increase the number of significantly overlapping troughs. As with the troughs applied to the convergence map, the troughs identified in the WL peak distribution trace different regions of the WL maps when GSN is added, and the degree of overlap between neighbouring troughs appears similar in both panels 1G and 2G.

Finally, Panels 1H and 2H show the SVF voids identified in the WL peak distribution. This algorithm produces the largest voids of all void finders and, similar to the WVF and tunnel algorithms, populates most of the area of the convergence map with voids. Also similar to the tunnels, the largest voids are in underdense regions and the smaller voids cluster in the overdense patches. It is interesting to note that in some cases, the tunnels and SVF identify the same voids in the WL peak distribution, as can be seen towards the top left of panels 1F and 1H. Panel 2H shows that the SVF voids identified in the WL peak distribution respond to GSN in the same way as tunnels and WVF, where these voids become smaller and more abundant in the presence of GSN.

\section{Void statistics} 
\label{sec:void statistics}

In this section we discuss the statistics of each of the seven void populations analysed here and study how the physical signal is affected by GSN in each case. We also investigate the impact of varying the smoothing scale to quantify how this mitigates the impact of GSN. For each void type we present the abundance, convergence profiles and tangential shear profiles. Then, in Section \ref{sec:comparison}, we will compare the different void populations and investigate which type of void is least affected by GSN while giving rise to the strongest tangential shear signature.

\subsection{Minima}
\label{sec:results:minima}

\begin{figure}
    \centering
    \includegraphics[width=\columnwidth]{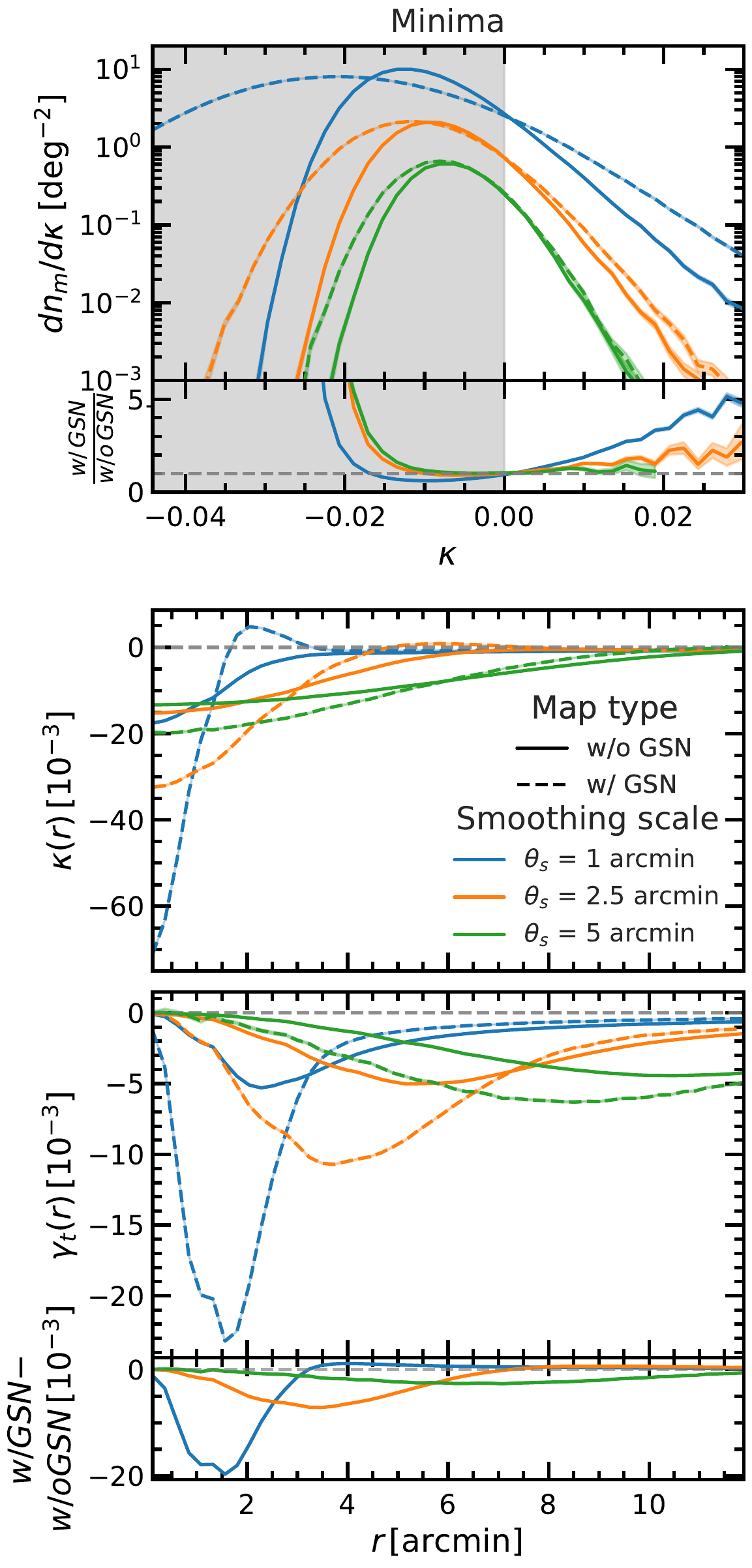}
    \vskip -.2cm
    \caption{The statistics describing the properties of WL minima depicted in panel E of Fig. \ref{fig:visualisation}. Solid lines show the properties of WL minima identified in WL maps with no GSN, while dashed lines show the properties of WL minima identified in WL maps with GSN. Different colours correspond to different smoothing scales applied to the convergence maps before identifying the minima, with blue, orange and green for $\theta_{\rm{s}}=1$, $2.5$ and $5$ arcmin respectively. One sigma standard error bars corresponding to the uncertainties associated to our analysis (which makes use of a 19200 deg$^2$ sky area) are given by the shaded coloured regions around each curve, however in most cases these error bars are a similar thickness to the curves. The top panel shows the WL minima abundance as a function of their WL convergence amplitude $\kappa$, and the shaded grey region indicates the minima that are used to calculate the lensing profiles. The middle panel shows the radial WL convergence profiles of the WL minima out to 12 arcmin, and the bottom panel shows the corresponding WL tangential shear profiles. The lower sub-panel in the top (bottom) panel shows the
    relative (absolute) difference between the minimum abundances (tangential shear profiles) measured in WL maps with and without GSN.
    }
    \label{fig:minima statistics}
\end{figure}

Fig.~\ref{fig:minima statistics} shows the statistics of the WL minima depicted in Panels 1E and 2E of Fig.~\ref{fig:visualisation} with and without GSN (dashed and solid lines respectively) for three smoothing scales, $1$, $2.5$ and $5$ arcmin (blue, orange and green respectively).

The top panel shows the differential WL minima abundance as a function of amplitude $\kappa$. The distribution peaks at $\kappa\sim-0.01$, with the peak shifting closer to $0$ as the smoothing scale increases. The distributions are also positively skewed, highlighting the non-Gaussian properties of WL minima. When GSN is included, the abundance of minima is significantly contaminated, especially for small smoothing scales. For $\theta_{\rm{s}} = 1$ arcmin, GSN introduces a large amount of spurious negative minima, while minima with such low negative amplitudes do not exist in the no GSN case. This is shown by the steep cutoff at $\kappa = -0.03$ for the no GSN case, while the minima abundance is still steadily decreasing below $\kappa = -0.03$ in the GSN-added case. A non negligible amount of spurious positive minima are also added by GSN, however this affect is less extreme than for negative minima. The creation of spurious minima due to GSN is suppressed as the smoothing scale increases, however even with $\theta_{\rm{s}} = 5$ arcmin, there is still a noticeable amount of spurious negative minima. For each smoothing scale it can be seen that the WL minima are significantly more impacted by GSN than WL peaks by comparing with the right panel of Fig.~\ref{fig:kappa pdf and WL peak abundance}.

Lensing profiles are calculated from minima with amplitudes $\kappa < 0$, as indicated by the shaded grey region in the top panel. The middle panel shows the mean stacked radial convergence profiles around the WL minima out to $12$ arcmin. For $\theta_{\rm{s}}=1$ arcmin, by comparing the blue solid and dashed lines, it can be seen that the addition of GSN artificially boosts the depth of the convergence profile at $r\sim0$ by over a factor of $3$. This is caused by the creation of a significant number of spurious minima with unphysically deep negative $\kappa$ values, as shown by the minima abundance in the top panel. For the GSN case, the minima convergence profile briefly becomes positive between $\sim1.5$ and $3$ arcmin, which is possibly due to the creation of spurious (negaitve) minima in local overdensities from GSN. In contrast, for the no GSN case, the convergence profile gradually approaches the mean background value of $\kappa = 0$. For larger smoothing scales, similar behaviour is still present, with the $\kappa$ amplitude at $r = 0 $ still artificially boosted by GSN, however this boost decreases with increasing smoothing scale.

The bottom panel shows the tangential shear profiles around the WL minima, $\gamma_{\rm{t}}(r)$, calculated from $\kappa(r)$ using Eq. \eqref{eq:gamma_t}. As the smoothing scale increases, the peak of the tangential shear profile moves to outer radii, whereas the inclusion of GSN shifts the peak to inner radii relative to the no GSN case. The difference in amplitude between the no GSN and GSN cases for the tangential shear profiles is smaller than for the convergence profiles, but significant contamination due to GSN still remains. For the no GSN maps, the height of the peak of the tangential shear profiles is somewhat insensitive to the smoothing scale, whilst increasing $\theta_{\rm{s}}$ quickly suppresses the peak in the tangential shear profiles for the GSN-added maps.

These statistics in Fig.~\ref{fig:minima statistics} show that the WL minima are significantly affected by GSN and are more susceptible to GSN than WL peaks.

\subsection{Troughs in the convergence map}
\label{sec:results:troughs_kappa}

\begin{figure*}
    \centering
    \includegraphics[width=2\columnwidth]{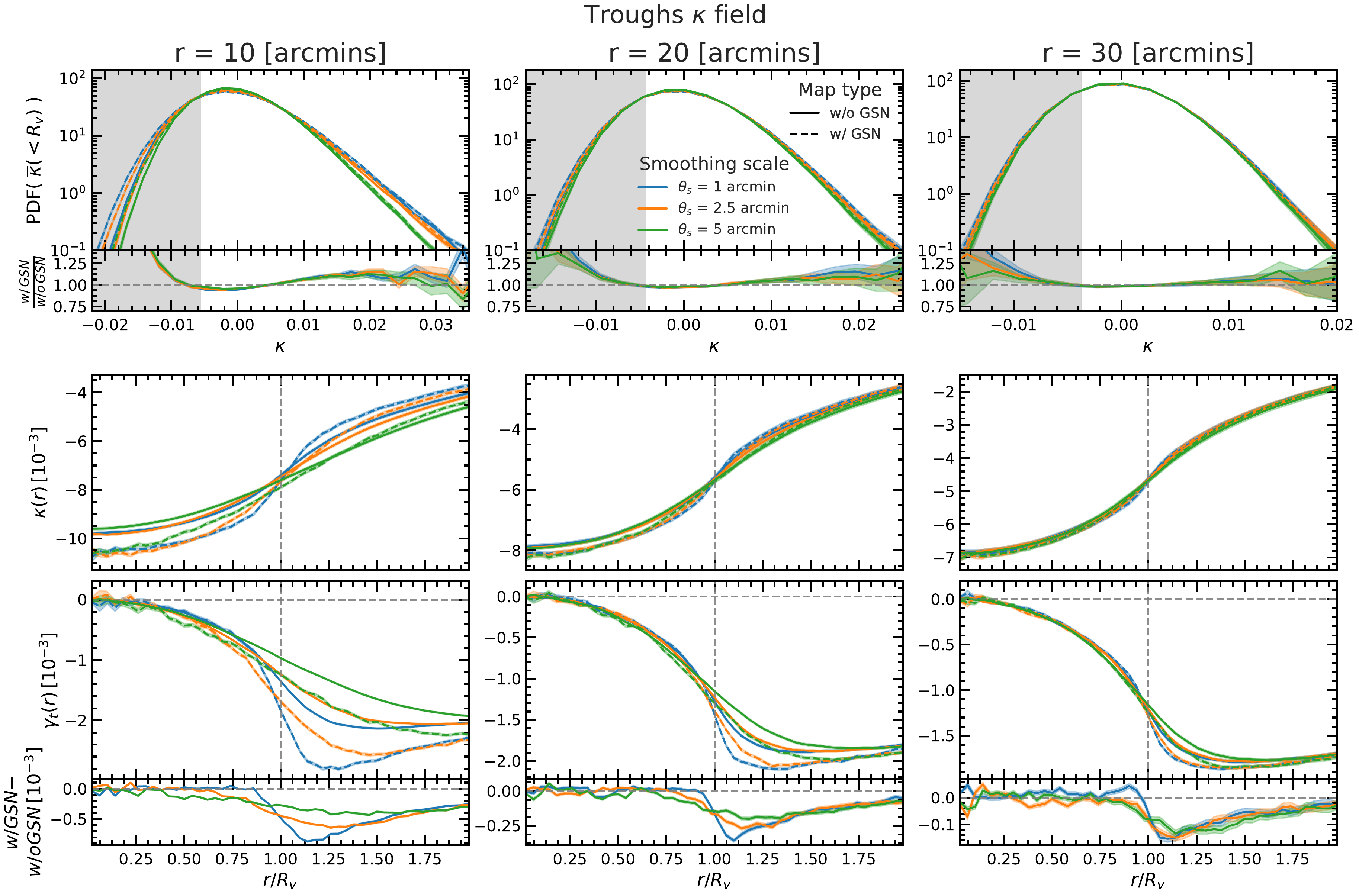}
    \caption{The statistics describing troughs identified directly in the convergence field. For the meaning of line colours and line types see the legend and, for more details, the caption of Figure \ref{fig:minima statistics}. The top row shows the PDF of the mean enclosed convergence within all randomly placed circles. The shaded grey region indicates the circles we define as troughs, that is the ones with a mean enclosed convergence in the bottom $20\%$ of all circles 
    (here we show the threshold for maps without GSN and for $\theta_{\rm{s}} = 2.5$ arcmin; the exact threshold depends slightly on smoothing scale and if GSN is included). The middle row shows the mean convergence profiles and the bottom row shows the mean tangential shear profiles. The three columns correspond to troughs of different sizes: $10$ (left), $20$ (centre) and $30$ (right) arcmin. The lower sub-panels in the top (bottom) row shows the
    relative (absolute) difference between the trough $\kappa$ PDFs (tangential shear profiles) measured in WL maps with and without GSN.}
    \label{fig:Trough ConvField statistics}
\end{figure*}

Fig.~\ref{fig:Trough ConvField statistics} shows the statistics of troughs identified directly in the convergence field. The top row shows the probability distribution function (PDF) of the mean enclosed convergence within all randomly placed circles, and the three columns (from left to right) are for trough radius $R_{\rm{v}}$ equal to $10$, $20$ and $30$ arcmin respectively. The shaded grey regions show the circles with a mean enclosed convergence in the bottom $20\%$ of all circles, which are the troughs that are used to calculate the trough lensing profiles. For a fixed trough radius, the $\kappa$ value above which circles are discarded depends on the smoothing scale and whether or not the WL maps includes GSN. For simplicity the shaded grey regions shown here are for $\theta_{\rm{s}} = 2.5$ arcmin in WL maps without GSN. 

Increasing the smoothing scale $\theta_{\rm{s}}$ decreases the width of the PDFs, and improves the agreement between the no GSN and GSN maps. As with the minima abundances, the largest differences between the no GSN and GSN maps are found at the negative-$\kappa$ regions of the PDF. As the trough radius increases, the agreement between the no GSN and GSN maps improves, and so does the agreement between different smoothing scales. These PDFs are all positively skewed indicating that the troughs identify more underdense regions than overdense regions. 

The middle row shows the mean stacked convergence profiles of the troughs for different radii. The troughs have very underdense centres, and $\kappa$ gradually increases with $r$. This increase gets sharper near $r = R_{\rm{v}}$ and then slows down further outside the trough radius. The depth of the convergence profiles is larger for the GSN maps, and the smoothing scale has a relatively small impact. As the trough radius increases, the overall depth of the convergence profiles decreases, however the shapes of the convergence profiles remain the same. The impact of GSN on the convergence profile decreases with $R_{\rm{v}}$, with the case $R_{\rm{v}}=30$ arcmin showing little difference between the GSN and no GSN cases.

The bottom row shows the tangential shear profiles of troughs, which are characterised by a maximum amplitude that is roughly an order of magnitude smaller than that of the WL minima. The inclusion of GSN has little impact on the trough tangential shear profiles for $r\lesssim R_{\rm{v}}$ (especially for the 20 and 30 arcmin troughs). At larger distances, GSN leads to an increase in tangential shear which persists even up to $r=2R_{\rm{v}}$. The difference between the maximum tangential shear amplitude for the no GSN and GSN maps is very small relative to the same feature in the WL minima. The difference between the no GSN and GSN maps is somewhat insensitive to the smoothing scale, and depends more strongly on the trough radius. As the trough radius increases, the amplitude of the tangential shear profiles decreases slightly and so does the difference between the no GSN and GSN maps. 

The statistics describing the troughs identified directly in the convergence maps are significantly less contaminated by the inclusion of GSN than the WL minima. However, the overall amplitude of the tangential shear profile of troughs is also significantly lower, which, as we shall see in Section~\ref{sec:comparison}, implies that we need a larger survey to measure trough profiles with the same SNR as the minima profiles.

\subsection{Troughs in the peak distribution}
\label{sec:results:troughs_peak}

\begin{figure*}
    \centering
    \includegraphics[width=2\columnwidth]{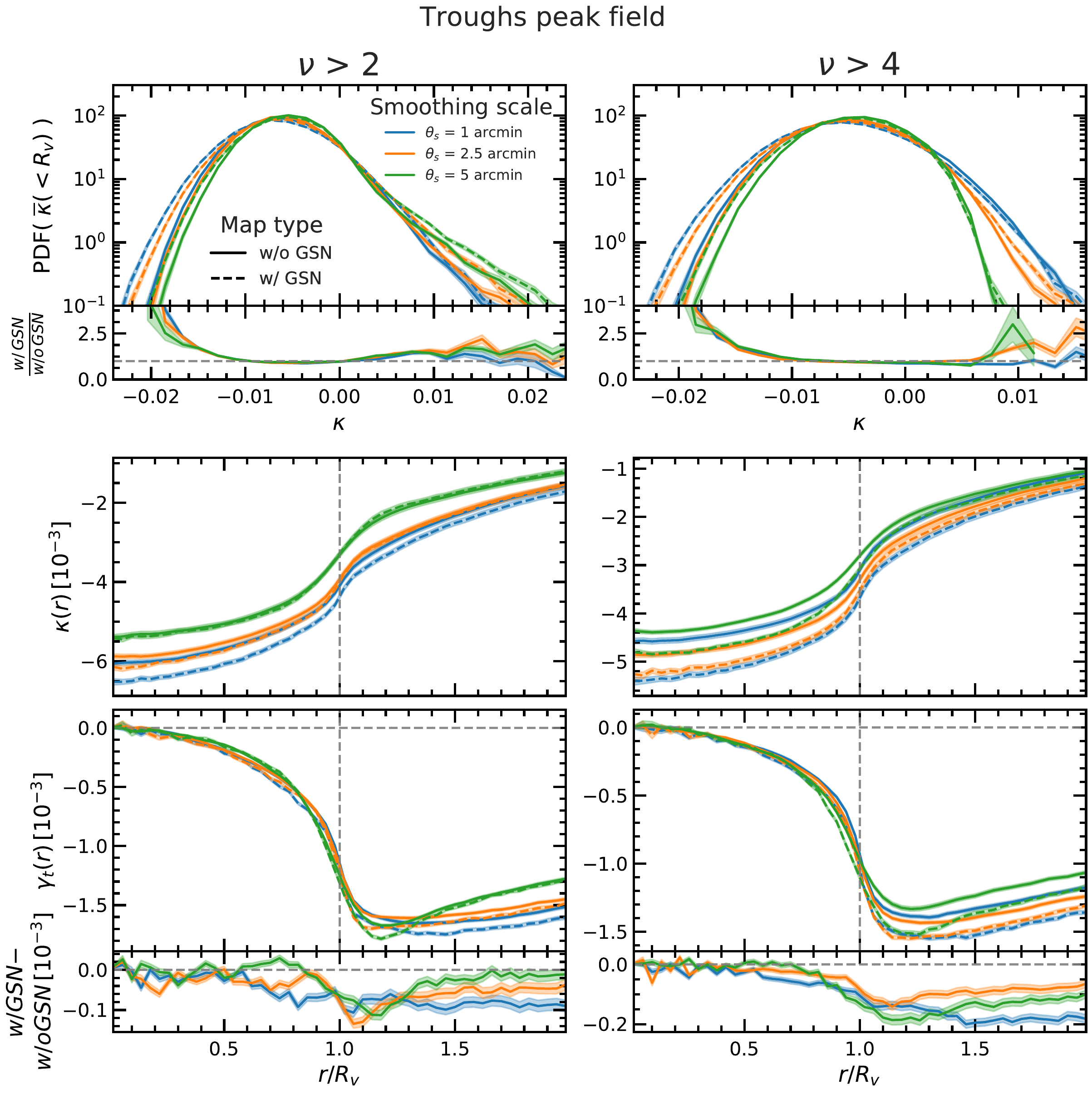}
    \caption{The statistics for troughs identified in the distribution of WL peaks. For the meanings of line colours and line types see the legend and, for more details, the caption of Figure \ref{fig:minima statistics}. The top row shows the PDF of the mean enclosed convergence within the troughs, the middle row shows the mean convergence profiles of the troughs and the bottom row shows the mean tangential shear profiles of the troughs. All results shown here are for a fixed trough size of $r = 30$ arcmin. We identify troughs using only the high WL peaks and we show results for two peak height selections: $\nu>2$ (left column) and $\nu>4$ (right column). The lower sub-panel in the top (bottom) panel shows the
    relative (absolute) difference between the trough $\kappa$ PDFs (tangential shear profiles) measured in WL maps with and without GSN.
    }
    \label{fig:Trough PeakField statistics}
\end{figure*}

We next study the troughs identified in the distribution of WL peaks. Before identifying troughs, we first remove all peaks below a predetermined $\nu$ threshold from the peak catalogue. This reduces the impact of GSN by discarding peaks with low height. This approach adds another free parameter to the void identification process compared to troughs identified in the convergence field, the $\nu$ threshold for peak heights. In Fig.~\ref{fig:Trough PeakField statistics}, we present results for two $\nu$ thresholds, $\nu > 2$ and $\nu > 4$, to test the impact of this threshold on the resulting trough statistics. To improve clarity, in Fig. \ref{fig:Trough PeakField statistics} all results are presented for a fixed trough size of $R_{\rm{v}} = 30$ arcmin, which is chosen because it is the trough radius at which results for the troughs agree best between the no GSN and GSN maps. 

The top row of Fig.~\ref{fig:Trough PeakField statistics} shows the PDFs of the mean enclosed convergence for troughs identified from WL peak catalogues with heights $\nu > 2$ and $\nu > 4$. Note that this is the trough PDF, which is calculated after the randomly placed circles with $\overline{\kappa}(<R_{\rm{v}})$ in the top $80\%$ are discarded, unlike in Fig.~\ref{fig:Trough ConvField statistics}. Away from the peak of the PDF, the results from the no GSN and GSN maps disagree for all smoothing scales for both peak thresholds. However, the agreement between the no GSN and GSN maps is good near the positive-$\kappa$ end of the PDF for all smoothing scales in the $\nu > 4$ catalogue. For the $\nu > 2$ catalogue, the PDFs are positively skewed, indicating that the trough algorithm is preferentially selecting underdense regions, however for the $\nu > 4$ catalogues the PDFs are more symmetric. This is due to the sparsity of tracers at this threshold, where the low number density of WL peaks implies that the $\nu > 4$ catalogue does not give an accurate representation of the underlying convergence field since, for example, many overdense regions of the convergence map do not have peaks with $\nu>4$. Despite this, the maximum of the PDF is still below zero indicating that we predominantly select underdense regions.

The middle row shows the radial convergence profiles of the troughs identified in the WL peak distribution. These profiles have a similar shape to those of the troughs identified in the WL convergence maps. For the $\nu > 2$ catalogue, agreement between the no GSN and GSN maps improves as the smoothing scale increases, and the two convergence profiles are within the one sigma standard error for $\theta_{\rm{s}} = 5$ arcmin. Here, the overall depth of the convergence profiles also decreases with increasing smoothing scale. However, for the $\nu > 4$ catalogues, increasing the smoothing scale only slightly improves the agreement between the no GSN and GSN maps, and there is no trend between smoothing scale and convergence profile depth, since $\theta_{\rm{s}} = 2.5$ arcmin produces the deepest convergence profile. This is due to the sparsity of WL peaks for $\nu > 4$, which results in the troughs more randomly tracing the underlying convergence field when compared to a lower $\nu$ threshold. This is evident from the fact that the convergence profiles are not as deep in the $\nu > 4$ catalogue when compared to the $\nu > 2$ catalogue.

The bottom row shows the radial tangential shear profiles of the troughs identified in the WL peak distribution. For all smoothing scales and both the no GSN and the GSN maps, the tangential shear profiles agree with each other reasonably well below $r = R_{\rm{v}}$, for both $\nu$ thresholds. This is due to the consistent shape of the convergence profiles (with only constant shifts with respect to each other) in all cases, which is the main feature that the tangential shear profile is sensitive to. The tangential shear profiles peak at $r \sim 1.2 R_{\rm{v}}$, which is where results from the different smoothing scales separate. The difference between the no GSN maps and the GSN maps is largest at the peak of the tangential shear, and slowly reduces out to larger radii. These tangential shear profiles are also noisier than for other void finders -- this is due to the larger scatter in the locations of the troughs identified in the peak distribution, as can be seen in Panel G of Fig.~\ref{fig:visualisation}, which results in a larger scatter of convergence profiles. 

Compared to troughs found directly in the $\kappa$ map, troughs identified using peaks have tangential shear profiles that have slightly lower amplitudes, however the agreement between the no GSN and GSN cases is better, which is a consequence of the fact that the WL peaks are less affected by GSN than the convergence field in the low $\kappa$ regions of the WL map.

\subsection{WVF voids}
\label{sec:results:WVF}

\begin{figure}
    \centering
    \includegraphics[width=\columnwidth]{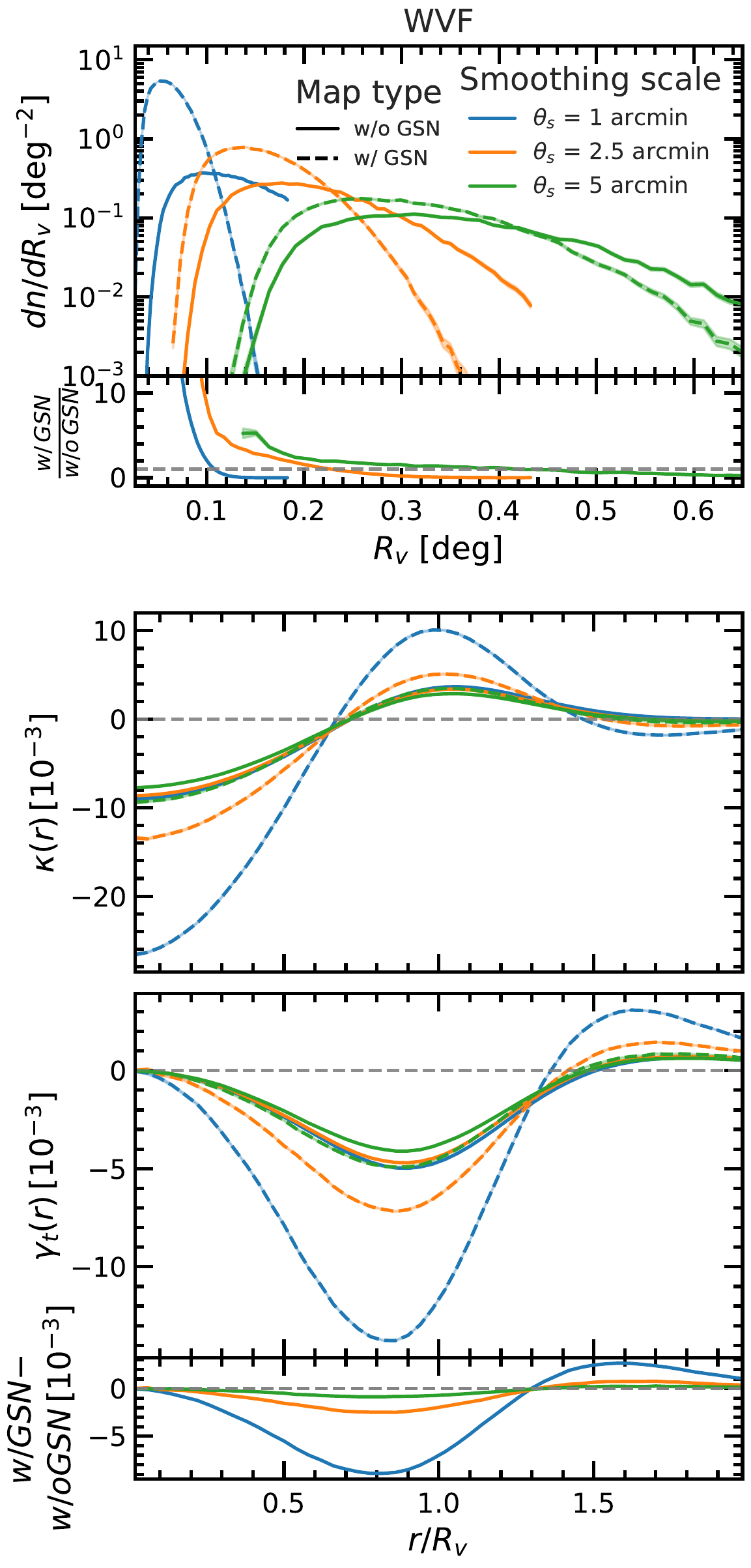}
    \caption{The abundance (top row), and the convergence (middle row) and tangential shear (bottom row) profiles of WVF voids. For the meanings of line colours and line types see the legend and, for more details, the caption of Figure \ref{fig:minima statistics}. The lower sub-panel in the top (bottom) panel shows the
    relative (absolute) difference between the WVF void abundances (tangential shear profiles) measured in WL maps with and without GSN.
    }
    \label{fig:WVF statistics}
\end{figure}

Fig.~\ref{fig:WVF statistics} shows the properties of the WVF voids. The top panel shows the differential void abundance as a function of void radius $R_{\rm{v}}$. For the smallest smoothing scale, the largest void that is identified is $0.2$ deg, and as the smoothing scale increases the sizes of the voids also increases, which also reduces the total number of voids. The size distributions of the voids are significantly different between the no GSN and GSN maps, where including GSN increases the total number of voids and reduces their size. This is due to GSN adding spurious features to the convergence field such as artificial ridges and minima, which results in the production of spurious voids. Since the WVF voids fill the entire area of the convergence map, having more voids implies that the average void size decreases. Even for $\theta_{\rm{s}} = 5$ arcmin, there is still a disagreement in the size distribution between the no GSN and GSN maps, and this disagreement is much larger than the one-sigma standard error bars (shown by the shaded regions around the curves).

The convergence profiles of WVF voids are shown in the middle panel. They have a smooth shape, with negative convergence values at $r = 0$, gradually increasing outwards and crossing $\kappa = 0$ at $r \sim 0.7R_{\rm{v}}$. The convergence profiles continue to smoothly increase until $r = R_{\rm{v}}$, at which point they start to decrease and return to the mean background value of $\kappa = 0$ far outside of the void radius. At $r \sim 1.5 R_{\rm{v}}$ some of the void profiles briefly become underdense, which is because the boundary of each void is also the boundary of one of its neighbours voids, which has an underdense interior. This feature is exaggerated for the smaller voids since averages are taken over smaller areas. 

In the absence of GSN, the convergence profiles are very similar for different $\theta_{\rm{s}}$ values. However, after adding GSN, the convergence profiles are heavily dependent of the chosen smoothing scale. For $\theta_{\rm{s}} = 1$ arcmin, the addition of GSN significantly reduces the $\kappa$ value at $r\sim0$, which is very similar to the behaviour seen in the WL minima convergence profiles. The similarity between the two is due to the fact that each watershed basin is connected to a local minima, which on average resides close to the centre of the void, and GSN produces a large number of spurious local minima, which can often be deeper than true minima (Fig.~\ref{fig:minima statistics}, top panel). This same feature will be seen in SVF voids found from the $\kappa$ field below. Furthermore, the amplitude of convergence profile in the positive regions is also boosted by GSN, which makes the peak at $r = R_{\rm{v}}$ significantly higher. The above behaviour occurs because the boundary of WVF voids consists of ridges in the $\kappa$ field and positive GSN values can move and enhance the height of these ridges (the algorithm chooses the highest local ridge and thus preferentially selects the regions with positive GSN values). This is more apparent for smaller smoothing scales, where GSN has not been sufficiently suppressed. The differences between the no GSN and GSN convergence profiles are quickly suppressed with increasing $\theta_{\rm{s}}$.

The bottom panel shows the tangential shear profiles for the watershed voids, which peak at $r\sim0.85R_{\rm{v}}$ and converge to $\gamma_{\rm{t}}\simeq0$ at large distances. Again, the $\gamma_{\rm{t}}$ profiles are significantly boosted by GSN, and quickly converge back to the no GSN counterparts as the smoothing scale increases. However, visible difference still remains even with $\theta_{\rm{s}}=5$ arcmin.

\subsection{SVF in the convergence map}
\label{sec:results:SVF_kappa}

\begin{figure}
    \centering
    \includegraphics[width=\columnwidth]{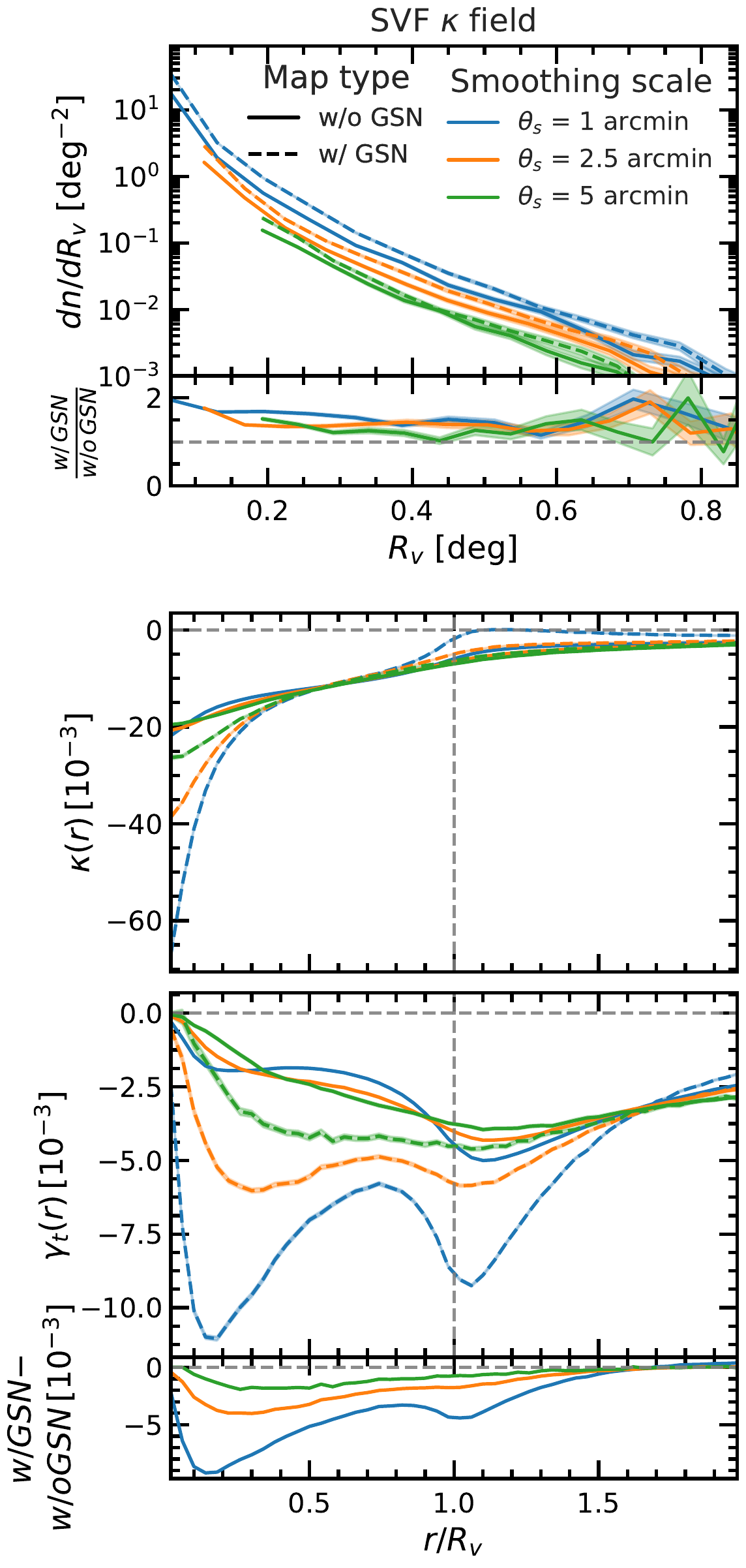}
    \caption{The statistics describing the SVF applied directly to the convergence maps: the abundance (top row), and the convergence (middle row) and tangential shear (bottom row) profiles of SVF $\kappa$ voids. For the meanings of line colours and line types see the legend and, for more details, the caption of Figure \ref{fig:minima statistics}. The lower sub-panel in the top (bottom) panel shows the
    relative (absolute) difference between the SVF-$\kappa$ void abundances (tangential shear profiles) measured in WL maps with and without GSN.
    }
    \vspace{-1em}
    \label{fig:SVF kappa statistics}
\end{figure}

Fig.~\ref{fig:SVF kappa statistics} shows the statistics for SVF voids identified directly in the convergence field (SVF $\kappa$). The shape of the void abundance function is different from the other void finders, declining faster with void radius than for other void types. Additionally, there is no turning point at the small-radius part of the distribution. For example, the WVF finds few very small voids, where the abundance of small voids briefly increases as the void radius increases, before the peak of the distribution. This is not the case for the abundance of SVF $\kappa$ voids, which does not reach a peak even at the smallest radii plotted. This is due to the SVF identifying voids with sizes down to the pixel resolution. As mentioned above, in this work we remove very small voids by imposing a minimum void size, $R_{\rm{v}} \geq 2\theta_{\rm{s}}$. 

The abundance of voids is systematically larger for the GSN maps than the no GSN maps, for all smoothing scales. In the case of the WVF, GSN increases the abundance of small voids but decreases the abundance of large voids, due to spurious structures introduced by GSN splitting the larger voids into smaller objects. For the SVF, the abundance of large voids is much lower to start with, and the voids populate the convergence maps much more sparsely, as shown in Panel D of Fig.~\ref{fig:visualisation}. This means that the spurious structures introduced by GSN contribute less to the degradation of true voids and largely only produce spurious voids, which is due to the addition of spurious minima from GSN (Fig. \ref{fig:minima statistics}, top panel) which are the seeds for the SVF $\kappa$ voids; this can be visibly seen by comparing panels 1D and 2D in Fig.~\ref{fig:visualisation}. Also, note that the abundance of SVF $\kappa$ voids decreases for all void radii when $\theta_{\rm{s}}$ increases, which is because the abundance of WL minima decreases with increasing $\theta_{\rm{s}}$, as shown by the top panel of Fig.~\ref{fig:minima statistics}.

The middle panel shows the mean radial convergence profiles of the SVF $\kappa$ voids. These voids are very deep at $r\sim0$, similar to the WL minima, and the convergence increases continuously out to $r=2R_{\rm{v}}$. Like in the WFV case, the convergence profiles in the no-GSN maps are somewhat insensitive to the chosen smoothing scale, whereas the depth of the profiles for the GSN maps is quickly suppressed with increasing $\theta_{\rm{s}}$. The depth of the convergence profiles at $r\sim0$ is artificially boosted when GSN is included (e.g. by a factor of 3 for $\theta_{\rm{s}}=1$ arcmin), which is again due to the creation of spurious minima with very low $\kappa$ values. However by $r=0.5R_{\rm{v}}$ the no GSN and GSN maps agree reasonably well, apart from the voids in the GSN added map for $\theta_{\rm{s}}=1$ arcmin, whose convergence profile returns to $\kappa=0$ faster than the other voids. 

The bottom panel shows the tangential shear profiles for the SVF $\kappa$ voids. For all other void finders, the inclusion of GSN boosts the amplitude of the tangential shear profile, and in some cases also changes slightly the radius where the signal reaches maximum. For the SVF $\kappa$ voids, the $\gamma_{\rm{t}}$ signal, which is maximal at $r\sim1.1R_{\rm{v}}$, is also boosted in the GSN maps relative to the no GSN maps. But here we find a secondary peak of $\gamma_{\rm{t}}$ at $r/R_{\rm{v}}\sim0.15$, which is particularly strong for small smoothing scales and when GSN is included. This is due to the flattening of the $\kappa$ profile at $0.3\lesssim r/R_{\rm{v}}\lesssim 0.8$ following a steep increase at $r/R_{\rm{v}}\lesssim0.3$. Such a large inner gradient of the $\kappa$ profile is due to these voids being centred on local WL minima, and this is more true in the GSN maps for which many of the SVF void centres correspond to spurious WL minima that are typically considerably deeper than the physical minima, as can be seen from the abundance of WL minima shown in the top panel of Fig.~\ref{fig:minima statistics} (and also the middle panel of Fig.~\ref{fig:minima statistics}). These spurious minima, on average, have much lower $\kappa$ values than their neighbours that manifests as a strong $\kappa$ gradient, which explains why the secondary peak is more pronounced for the case of GSN maps.

The agreement between the tangential shear profiles in the no-GSN maps and the GSN maps improves slightly as the smoothing scale increases. However, a significant difference remains even for $\theta_{\rm{s}}=5$ arcmin, as in the case of WVF voids, highlighting the fact that the impact of GSN is hard to be completely eliminated for voids identified from the WL convergence map.

\subsection{SVF in the peak distribution}
\label{sec:results:SVF_peak}

\begin{figure*}
    \centering
    \includegraphics[width=2\columnwidth]{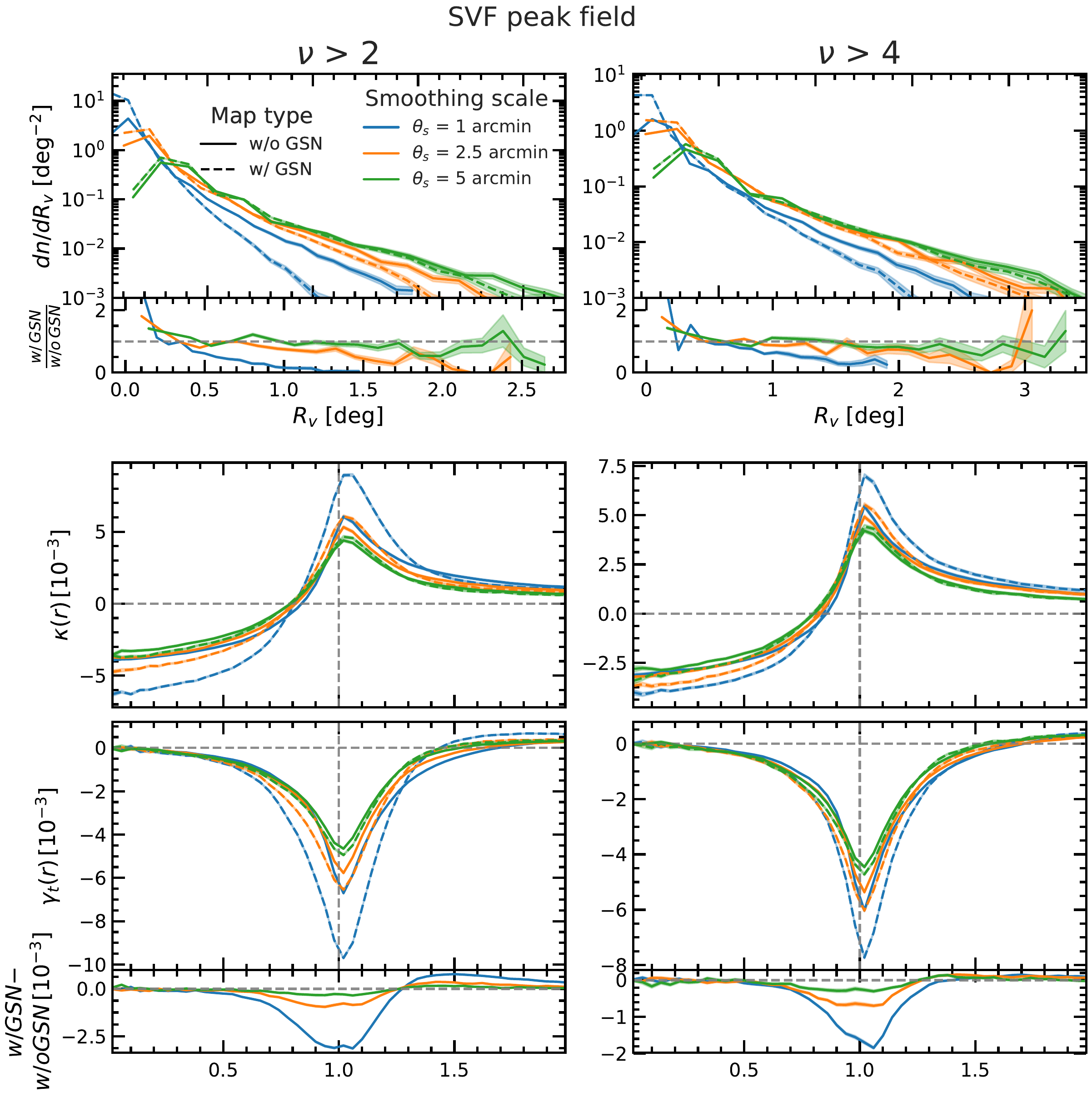}
    \caption{The statistics describing the SVF applied to the WL peak distribution: the abundance (top row), and the convergence (middle row) and tangential shear (bottom row) profiles of SVF peak voids. For the meanings of line colours and line types see the legend and, for more details, the caption of Figure \ref{fig:minima statistics}. Each column corresponds to voids identified in a different WL peak catalogue, $\nu>2$ on the left and $\nu>4$ on the right. The lower sub-panel in the top (bottom) panel shows the relative (absolute) difference between the SVF-peak void abundances (tangential shear profiles) measured in WL maps with and without GSN.
    }
    \label{fig:SVF peak statistics}
\end{figure*}

Fig.~\ref{fig:SVF peak statistics} shows the statistics for SVF voids identified in the WL peak distribution (SVF peak). The top panel shows the differential void abundance. The SVF peak algorithm identifies the largest voids of all the void finders studied in this work, with some voids as large as two degrees in radius. Here larger smoothing scales reduces the total number of voids but creates larger voids, and including GSN adds spurious small voids and reduces the abundance of large voids. This is due to the generation of spurious WL peaks from the addition of GSN, where a higher number density of tracers split large voids into multiple smaller ones. Fewer voids are detected overall in the $\nu > 4$ catalogue compared to the $\nu > 2$ catalogue, however these voids are larger than their counterparts in the $\nu > 2$ catalogue. This is again due to the reduced number density of WL peaks that are used as tracers in the void identification. Apart from this the abundances of the voids in the two catalogues appear qualitatively similar. 

The middle row shows the convergence profile for the SVF peak voids, which are underdense close to the void centre and overdense near the void boundary. Outside of the void radius the convergence gradually approaches the background value of $\kappa=0$. The depths of the void centres and amplitudes at the void radius are boosted in the GSN maps, however the difference between the void convergence profiles in the no-GSN and GSN added maps is quickly suppressed as the smoothing scale increases, and at $\theta_{\rm{s}}=5$ arcmin the difference is small. The depth close to the void centres and the peak at the void boundary also decrease when the smoothing scale increases. These voids are less underdense than most of the other void types. 

The bottom row presents the tangential shear profiles for the SVF peak voids. These profiles have a sharp peak at $r = R_{\rm{v}}$ and the amplitude of these peaks is large despite the shallow convergence profiles near the void centres. This is due to the rapid increase in $\kappa(r)$ seen in the range $r/R_{\rm{v}}\in[0.7,1.0]$, with the $\gamma_{\rm{t}}(r)$ amplitude being largest when $\kappa(r)$ changes rapidly. This highlights that identifying the deepest underdensities is not the most important criteria when the tangential shear profile is the observable of main interest. Similar to the other void finders, the peak of the tangential shear profiles is boosted in the GSN maps, however, as with the convergence profiles this difference is quickly suppressed as $\theta_{\rm{s}}$ increases, with most of the difference removed with $\theta_{\rm{s}} = 5$ arcmin. The amplitude of the tangential shear profiles is slightly smaller for the peak catalogue with a larger $\nu$ threshold, indicating that it does not depend strongly on the $\nu$ threshold used for WL peak selection. The main difference comes from the fact that having a higher $\nu$ threshold results in fewer voids that, as we shall see in Section~\ref{sec:comparison}, means a lower SNR when measuring the shear profiles of these voids for a given sky area.

\subsection{Tunnels}
\label{sec:results:tunnels}

\begin{figure*}
    \centering
    \includegraphics[width=2\columnwidth]{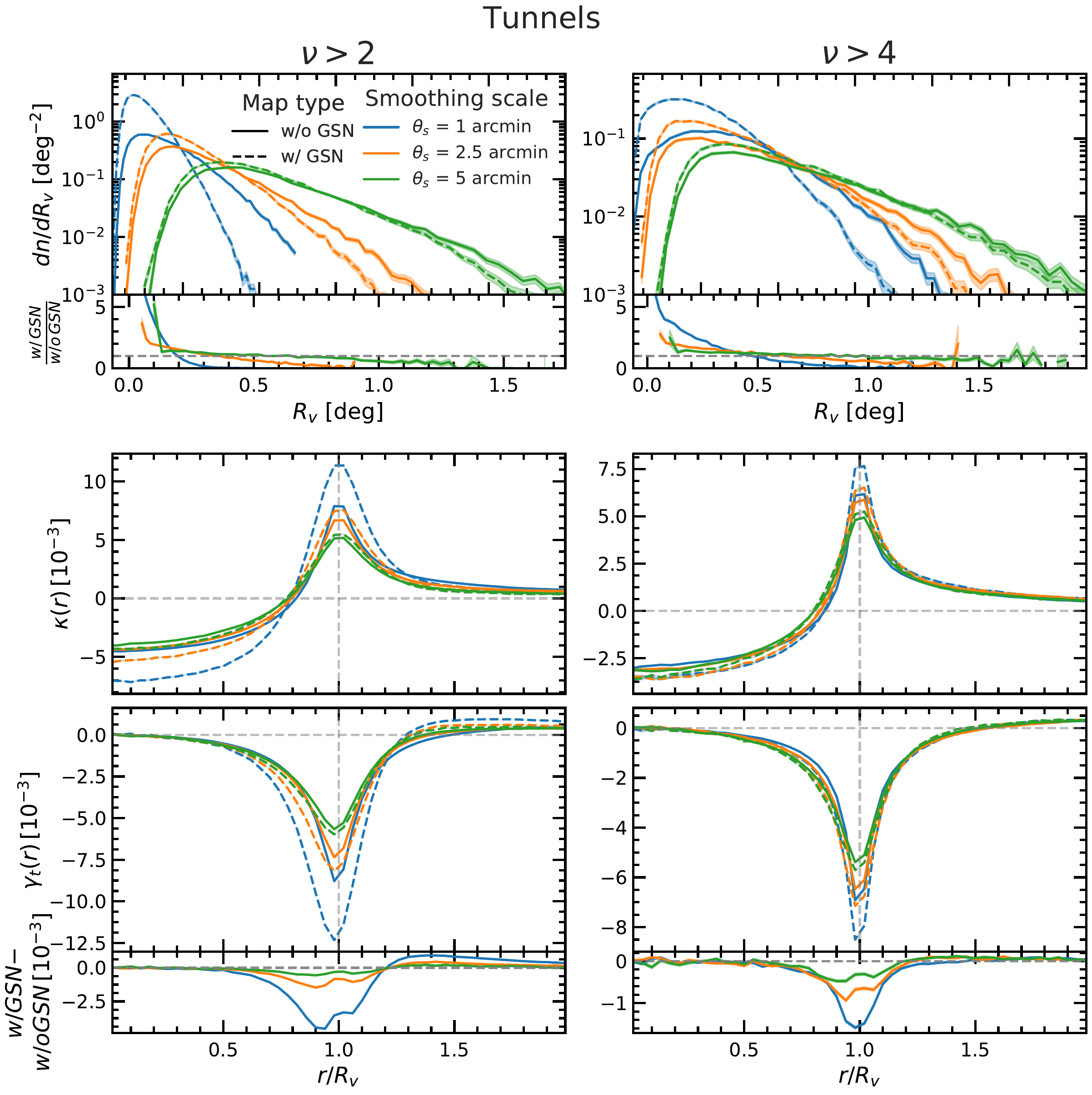}
    \caption{The statistics describing the Tunnels identified in the WL peak distribution: the abundance (top row), and the convergence (middle row) and tangential shear (bottom row) profiles of tunnels. For the meanings of line colours and line types see the legend and, for more details, the caption of Figure \ref{fig:minima statistics}. The left and right columns correspond to tunnels identified in WL peak catalogues with heights $\nu >2$ and $\nu > 4$ respectively. The lower sub-panel in the top (bottom) panel shows the relative (absolute) difference between the tunnel abundances (tangential shear profiles) measured in WL maps with and without GSN.}
    \label{fig:tunnels statistics}
\end{figure*}

Fig.~\ref{fig:tunnels statistics} shows the statistics of voids identified in the WL peak distribution using the tunnel algorithm, where the left and right columns correspond to tunnels identified in WL peak catalogues with heights $\nu > 2$ and $\nu > 4$ respectively. The top row shows the differential void abundance of the tunnels. The tunnel algorithm also identifies some of the largest voids studied in this work, although the largest SVF peak voids are larger than the largest tunnels. Consistent with other void finders, the tunnel algorithm identifies more voids in total in the maps that include GSN, and fewer large voids. The abundance of the tunnels decreases, and the size of the tunnels, increases with increasing $\theta_{\rm{s}}$. The differences in the void abundances between the no-GSN and GSN maps decreases with increasing $\theta_{\rm{s}}$ and the difference becomes small at $\theta_{\rm{s}}=5$ arcmin.

The middle row shows the tunnel convergence profiles, which have a very similar shape to that of the SVF peak voids. This is to be expected as in some cases both of these algorithms identify the same voids. Beyond their similarities, the tunnel algorithm identifies voids with slightly deeper convergence profiles near the centre and more overdense ridges at the boundary. This is because the tunnels by definition do not enclose any WL peaks but instead only have peaks residing at their boundaries, whereas the SVF peak algorithm allows WL peaks to reside within voids, which can lead to higher $\kappa$ values inside SVF peak voids than inside tunnels. Similar to other void types, adding GSN leads to lower $\kappa$ values at the tunnel centres and a higher overdensity at the tunnel boundaries. This difference is again strongly suppressed for $\theta_{\rm{s}}=5$ arcmin. The tunnels behave similarly to the SVF peak voids when the $\nu$ threshold of the WL peak catalogue is increased, slightly reducing the depth of $\kappa$ profiles at the void centre and the peak at the void boundary, whilst the peak becomes sharper. 

The bottom row shows the tangential shear profiles which are qualitatively similar to the results of SVF peak voids, except that the tunnels have a higher peak at $r = R_{\rm{v}}$. The difference between the no-GSN and GSN-added maps respond to the chosen smoothing scale in the same way as the convergence profile, with little difference remaining when $\theta_{\rm{s}}$ increases to $5$ arcmin. Changes in the tangential shear in response to increasing the $\nu$ threshold are also the same as in the convergence profiles. Here we note that for the $\nu > 4$ WL peak catalogue, the convergence and tangential shear profiles for all smoothing scales, and for maps with and without GSN, are all very similar and follow each other closely, overlapping in some places. The main difference between the different curves can be seen at the peak of the profiles where most of the information in terms of SNR is contained \citep{Cautun2018}.

\section{Comparison of different void definitions}
\label{sec:comparison}

In this section we quantify the relative merit of each void finder. There are many criteria that one could use to quantify the suitability of a specific void finder for a given purpose \citep[e.g. see][]{Cautun2018,Paillas2019}. Here we are interested in a rather general comparison of the various methods that identify WL voids. We choose to do so by answering two questions: i) Which void populations are least affected by GSN? and ii) Which void types have the highest tangential shear signal, as quantified in terms of SNR? These questions are motivated by the goal of using WL voids to constrain cosmological parameters and alternative cosmological models. To a first approximation, we expect that the constraints derived from voids will be maximal when their signal, such as $\gamma_{\rm{t}}$ profiles, can be measured with low uncertainties (i.e., high SNR) and when the effects of GSN are minimised \citep[e.g. see][]{Cautun2018,Paillas2019}. This might not always be the case as we discuss later on, but nonetheless is a good starting point for a general comparison.

\subsection{Impact of GSN}
\label{sec:comparison:GSN}

GSN is the leading contribution to noise that contaminates the observed WL signal, and for this reason it is important to understand how the void finders respond to GSN, before the statistics developed here can be used to constrain cosmological parameters. As we saw in Section~\ref{sec:void statistics}, GSN can lead to the identification of spurious voids and to the breaking of physical voids into more objects. This could potentially degrade the cosmological information contained in the statistics of voids, and thus lower the cosmological constraints that can be inferred using WL voids. 

To assess the effect of GSN, we proceed by comparing voids in maps with and without GSN. Such a test requires us to choose a WL void statistic to measure the impact of GSN. Up to now, we have studied the abundances and $\gamma_{\rm{t}}$ profiles with and without GSN, and here we choose to focus on the tangential shear profile, which has been shown to provide tighter cosmological constrains, such as when testing modified gravity models \citep[e.g.][]{Davies2019b}. We measure the change in the amplitude of the $\gamma_{\rm{t}}$ signal when GSN is added, as a means to quantify the impact of GSN on the lensing profile. Typically, for the void $\gamma_{\rm{t}}$ profiles most of the cosmological constraining power comes from the bins where the amplitude of the signal is maximal \citep[e.g.,][]{Cai2015,Barreira2015,Cautun2018,Davies2019b} and, as such, we measure the impact of GSN at this location.

\begin{figure*}
    \centering
    \includegraphics[width=2\columnwidth]{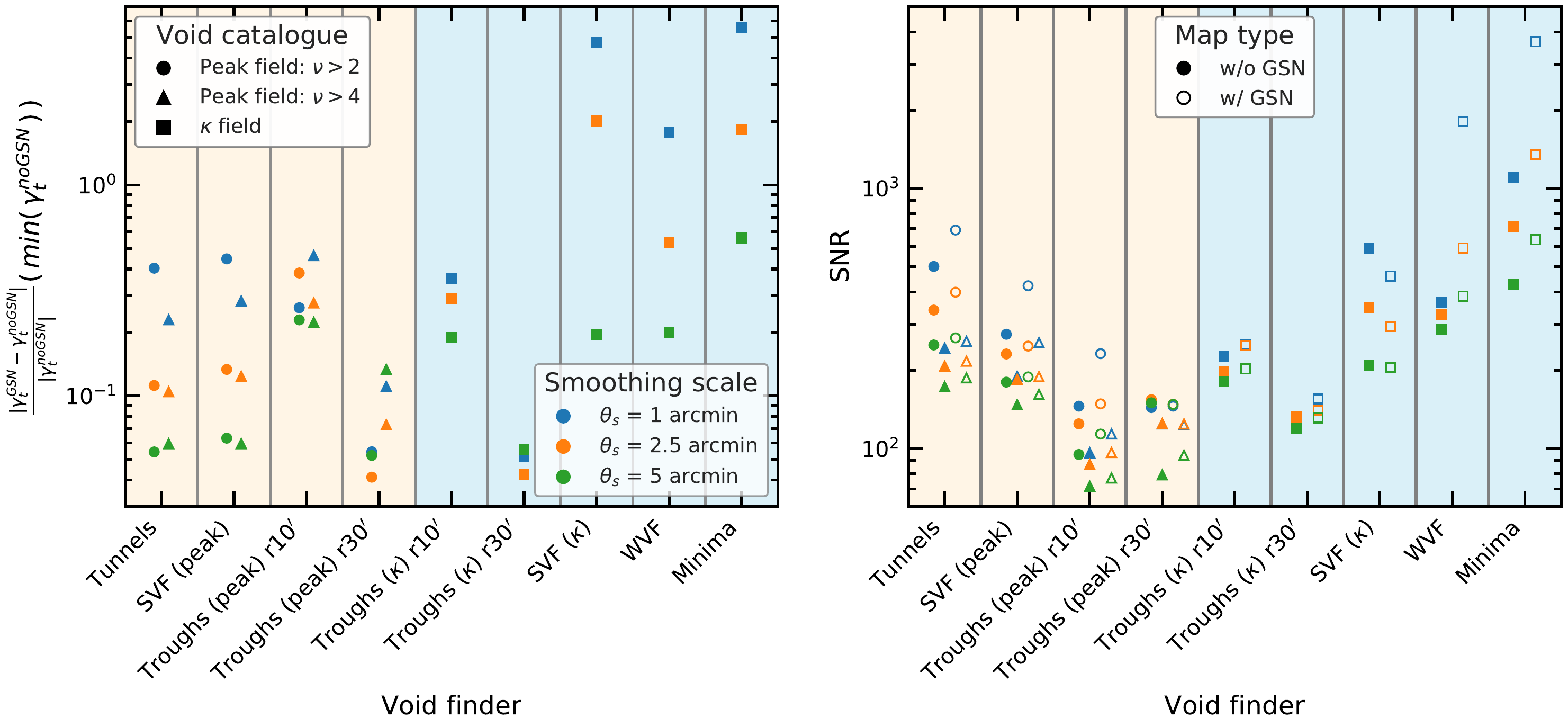}
    \caption{Comparisons of the seven void populations studied here in terms of the impact of GSN and in terms of the SNR associated to the tangential shear measurement for a \LSST{} like survey.
    \textit{Left panel:} the relative difference between $\gamma_{\rm{t}}$ in the GSN-added and no-GSN convergence maps, at the radius at which the amplitude of $\gamma_{\rm{t}}$ in the no-GSN maps is highest ($\gamma_{\rm{t}}$ is lowest).
    \textit{Right panel:} An $\LSST$ forecast of the total SNR with which the $\gamma_{\rm{t}}(r)$ profile will be measured for each void type. All results in both panels are for all void finders studied in this work (x-axis). 
    A yellow background indicates results for void finders applied to the WL peak distribution and a blue background indicates results for void finders applied directly to the WL convergence maps. Circles correspond to results from voids identified in WL peak catalogues with $\nu > 2$, triangles are for $\nu > 4$, and squares are from voids identified directly in convergence maps. Blue, orange and green markers indicate
    different smoothing scales, with $\theta_{\rm{s}} = 1, 2.5$ and $5$ arcmin, respectively. In the right panel solid markers indicate results from no GSN maps, and empty markers show results for WL maps with GSN added. Here we plot troughs with a radius of $R_{\rm{v}}=10$ and $30$ arcmin (labelled as $r10^\prime$ and $r30^\prime$, respectively), to show the impact of changing the trough radius.
    }
    \label{fig:diff_and_SNR}
\end{figure*}

The left panel of Fig.~\ref{fig:diff_and_SNR} shows the relative difference, $|\gamma_{\rm{t}}^{\rm GSN} - \gamma_{\rm{t}}^{\rm no-GSN}|/ |\gamma_{\rm{t}}^{\rm no-GSN}|$, between $\gamma_{\rm{t}}$ in the GSN-added and no-GSN convergence maps, at the radius at which the amplitude of $\gamma_{\rm{t}}$ in the no-GSN is maximal (i.e., where $\gamma_{\rm{t}}$ has the most negative value), for all void finders studied in this work. Here, lower values correspond to a small relative impact on the $\gamma_{\rm{t}}$ amplitude from GSN while large values indicate that GSN is significantly boosting the $\gamma_{\rm{t}}$ amplitude (for all void populations studied here, GSN always increases the amplitude of the $\gamma_{\rm{t}}$ signal; see Appendix~\ref{app:WL voids in GSN maps} for a discussion of the reason behind that).

We find that GSN has the largest impact on the $\gamma_{\rm{t}}$ profiles of WL minima. This is due to the fact that GSN creates more spurious minima than spurious structures in the other void finders, which is one drawback of the simplicity of the WL minima definition. The boost from GSN is somewhat decreased for the minima when larger smoothing scales are applied. However, in many cases the boost to the minima $\gamma_{\rm{t}}$ profiles from GSN with $\theta_{\rm{s}} = 5$ arcmin (about $55\%$) is larger than the $\gamma_{\rm{t}}$ boost from GSN for other void finders with $\theta_{\rm{s}}=1$ arcmin. The $\gamma_{\rm{t}}$ signal for SVF $\kappa$ is also boosted by GSN by a similar (relative) amount as the WL minima, which is due to the minima being used as prospective void centres at the start of the SVF $\kappa$ void identification process. For SVF $\kappa$ the relative difference between the no-GSN and GSN $\gamma_{\rm{t}}$ amplitudes is more quickly suppressed by increasing $\theta_{\rm{s}}$ than for the WL minima, reaching $\sim20\%$ for $\theta_{\rm{s}}=5$ arcmin. The WVF voids also appear to respond to GSN in a similar way to the WL minima and SVF $\kappa$, however the amplitude of the boost due to GSN is slightly lower. Finally, for all of the void finders applied directly to the convergence maps, troughs $\kappa$ appears to be the least impacted by GSN, and they also see the smallest impact on the agreement between the no-GSN and GSN maps from increasing $\theta_{\rm{s}}$, as can also be seen in Fig.~\ref{fig:Trough ConvField statistics}. 

The void populations that are the least impacted by GSN are those identified in the distribution of WL peaks. This is due to high amplitude WL peaks (Fig.~\ref{fig:kappa pdf and WL peak abundance}, right panel) being more resilient to GSN than underdense regions, i.e., $\kappa<0$, which are the ones determining most of the properties of voids identified directly in the convergence field.

We find that both the tunnels and SVF peak voids respond to GSN in very similar ways and that the impact of GSN is reduced for voids identified in peak catalogues with larger $\nu$ thresholds. Finally, the trough peak void finder is the most resilient to GSN of all the methods that employ WL peaks, however in contrast to the tunnels and SVF peak, the impact of GSN increases when the $\nu$ threshold increases, which is because troughs peak is more sensitive to tracer sparsity than tunnels and SVF peak.

Both of the trough algorithms are the least impacted by GSN for $R_{\rm v} = 30$ arcmin, however for a trough radius of $10$ arcmin, the impacts of GSN on the tangential shear profiles for both trough peak and trough $\kappa$ voids becomes worse than tunnels and SVF peak. 

\subsection{The SNR of tangential shear profiles}
\label{sec:comparison:SNR}

Next we investigate the signal-to-noise ratio (SNR) with which we can measure the tangential shear signal of WL voids. Our goal is to assess which void type has the largest SNR since potentially those voids are the most promising to use for cosmological constraints. For examples, \citet{Cautun2018} and \citet{Paillas2019} have studied the signature of modified gravity models in the void population identified using multiple void finders. For 2D voids, they have found that all methods show roughly equal fractional differences in the void shear profiles when comparing modified gravity with the standard model, and thus the optimal void type to constrain such alternative cosmological models is the one in which the $\gamma_{\rm{t}}$ profile can be measured with the highest SNR.

We define the SNR with which we can measure the tangential shear profile of voids as:
\begin{equation}
    {\rm{SNR}}^2 \equiv \sum_{i,j} \gamma_{\rm{t}}(i) \,\, \alpha \,\,  {{\rm Cov}}^{-1}(i,j) \,\, \gamma_{\rm{t}}(j)
    \label{eq:SNR} \;,
\end{equation}
where the sum is over all bins of $r/R_{\rm{v}}\in[0,2]$, $i$ and $j$ denote the bins to be summed over, and ${\rm Cov}^{-1}$ is the inverse of the covariance matrix for the tangential shear measurements. Here $\gamma_{\rm{t}}$ is the mean tangential shear measured from all voids from all 192 maps used in this study and $\alpha$ is the Anderson-Hartlap factor \citep{Anderson2003, Hartlap2007} which we use to compensate for the bias introduced by inverting a noisy covariance matrix. The $\alpha$ factor is given by
\begin{equation}
    \alpha = \frac{ N - N_{\rm{bin}} - 2 }{N - 1}
    \; ,
\end{equation}
where $N = 192$ is the number of realisations used to calculate the covariance matrix, and $N_{\rm{bin}} = 50$ is the number of radial bins. We calculate the covariance matrix using the central  \map{10} region of the 192 maps described in Section~\ref{sec:Weak lensing maps}. We then rescale the SNR values by $\sqrt{ A_{ \rm{LSST} } / {A} } = 13.4$ in order to present a forecast for an \LSST{} like survey that has a sky coverage, $A_{ \rm{LSST}} = 18,000$ deg$^2$. 

The right panel of Fig.~\ref{fig:diff_and_SNR} shows the SNR (see Eq.~\eqref{eq:SNR}) for the tangential shear profiles from each void finder we have studied. The coloured symbols indicate the results for the three smoothing scales we have studied and we present the SNR values for convergence maps with (open symbols) and without (filled symbols) GSN. This allows us to characterise how the SNR changes when identifying voids in noisy maps.

For all void types, we find that increasing the $\theta_{\rm{s}}$ smoothing length decreases the SNR ratio; the only exceptions are the troughs peak ($R_{\rm{v}} = 30$ arcmin) and troughs $\kappa$ voids ($R_{\rm{v}} = 10$ and $30$ arcmin), for which the SNR is roughly the same for all three smoothing scales that we used. For the voids found in the peak distribution, increasing the peak threshold leads to lower SNR. Thus, the SNR is maximised for small smoothing scales and for peak catalogues with small $\nu$ thresholds.

The right panel of Fig.~\ref{fig:diff_and_SNR} reveals a rather interesting result, which is surprising at first. All void types (except SVF $\kappa$) identified in the maps with GSN show a larger SNR than the voids found in the map without GSN. This might be counter-intuitive since, as we discussed, GSN fragments large voids into two or more components and adds spurious objects to the sample, which potentially reduces the sensitivity of voids to cosmology. The answer is given by the fact that the SNR we calculate describes how well we can measure the $\gamma_{\rm{t}}$ signal of a void and not the amount of cosmological information it contains.

The SNR of WL voids in maps with GSN is higher than for the maps without GSN due to two factors: i) adding GSN increases the amplitude of the mean $\gamma_{\rm{t}}$ profile, and ii) it leads to identifying more voids, as shown in Figs.~\ref{fig:minima statistics}-\ref{fig:tunnels statistics}. The change in void shear profiles and abundance is an artificial one and it is due to using the same noisy map to identify voids and calculate their profiles. For example, adding a negative GSN value to a pixel makes it more likely to be associated to the interior of a void, and, as a result, the interior of voids is deeper for maps with GSN since it is more likely to contain regions with negative GSN contributions than positive ones. The opposite holds true for the void boundaries. A pixel with a positive GSN value is more likely to be identified as part of a void's edge, and thus the void boundaries in maps with GSN contain a higher fraction of pixels with positive GSN values, which artificially boosts the mean $\kappa$ value at the void boundary. These two effects lead to an artificially stronger tangential shear profile for voids in GSN maps (for a more detailed discussion and examples see Appendix~\ref{app:WL voids in GSN maps}). 

We find that the WL minima tangential shear profiles have the largest SNR both in the no-GSN maps and the GSN-added maps, which indicates that they are promising cosmological probes. The WVF has the second highest SNR in the GSN-added maps, but is beaten by SVF $\kappa$ in the no-GSN maps. Both of the trough algorithms give the lowest SNR values despite being the least affected by GSN in the left panel of Fig.~\ref{fig:diff_and_SNR}. SVF peak gives reasonable SNR values, but fares slightly less well in almost all cases than tunnels, which gives SNR values comparable to the void finders applied directly to the WL convergence maps.

\subsection{Which void definition is best?}
\label{sec:comparison:best_void}

Ideally, the optimal void finder would be the one least affected by GSN while having the largest SNR for its tangential shear profile. Fig.~\ref{fig:diff_and_SNR} shows that these two requirements are not compatible: the void finders least affected by GSN (either troughs peak or troughs $\kappa$) have the lowest SNR for $\gamma_{\rm{t}}$, while the voids with the highest SNR (WL minima) are strongly impacted by GSN. The same behaviour is seen when varying the void parameters studied here. Increasing the $\kappa$ smoothing length, $\theta_{\rm{s}}$, used to identify voids, while lowering the impact of GSN, also decreases the SNR for tangential shear. For voids identified in the peak distribution, increasing the $\nu$ threshold used for selecting the peak catalogue mitigates the effect of GSN, but again reduces the $\gamma_{\rm{t}}$ SNR. Therefore, there is no clear choice for the best void finder or the best selection of void finding parameters, such as $\theta_{\rm{s}}$ or WL peak $\nu$ threshold.

In general, we find that the void finders that use WL peaks as tracers are less impacted by GSN, while the void finders applied directly to the WL convergence maps give higher SNR values. The void finder that generally offers a good compromise between minimal impact from GSN and a high SNR value is the tunnel algorithm. It has a $\gamma_{\rm{t}}$ SNR similar to that of the SVF and WVF $\kappa$ field voids finders while being the second least affected by GSN, after troughs.

We would also like to point out that GSN does not necessarily decrease the amount of cosmological information contained by a probe, and that in some special circumstances it can help make this information more easily accessible. For example, this has been pointed out by \citet{Yang2011}, who have shown that the abundance of WL peaks in maps that include GSN provides better cosmological constraints than for maps without GSN. \citeauthor{Yang2011} have attributed this effect to stochastic resonance, which is a well-studied phenomena \citep{Gammaitoni1998} where a signal in a physical system may be boosted when a source of noise is added, under certain conditions. The conditions required for stochastic resonance to take place within a system are: i) a form of a threshold, ii) a weak coherent input, and iii) a source of noise that adds to the coherent input. From the above it is clear that all three of these conditions apply to WL peaks as discussed in \citeauthor{Yang2011}, and hence they also apply to WL voids. The first requirement for stochastic resonance is a form of threshold, which in the context of WL voids is the criteria that all void finders identify underdense regions through one means or another. The second requirement is a weak coherent input, which in this context is the WL convergence map. The WL map can be considered weakly coherent because GSN dominates the signal (before smoothing), but contains coherent information due to physical correlations in the map induced by gravitational collapse. Finally, for stochastic resonance we require a source of noise that is added to the WL convergence map, which exactly matches our prescription for modelling GSN. 

In the case of WL voids, stochastic resonance occurs because the void finders are designed to identify underdense regions, or underdense regions enclosed by overdense regions etc.. The inclusion of GSN exaggerates some underdense regions and some overdense regions. However, since GSN is random and uncorrelated (neglecting higher order effects such as intrinsic alignment), it could also make some underdense and overdense regions flatter (i.e., smoothed out). Because all void finders fulfil a set of criteria when identifying voids, they will preferentially select the regions that have been exaggerated by GSN and neglect the regions that have been flattened by GSN. Furthermore, distinct deep voids in the physical maps (without GSN) are less likely to be removed by GSN, because the physical signal will dominate the GSN. However less distinct voids that might be missed in the physical maps have a chance to be randomly boosted by GSN, which will result in their detection in the GSN-added maps. These are competing factors with the consequence that GSN can affect true voids and generate spurious fake voids, though true voids are rarely destroyed by GSN but instead are most commonly split up into smaller voids (e.g., as discussed with the tunnel algorithm). It is currently unclear whether or not the boost in SNR from GSN seen in Fig.~\ref{fig:diff_and_SNR} will translate to improved parameter constraints relative to the case without GSN (which is unobservable), however we leave this to a future study. For this reason, we have focused on identifying the void finder that is the least impacted by GSN, whilst still producing high SNR values.

\section{Discussion and conclusions}\label{sec:discussion and conclusions}

In this paper we have presented a comparison of different void finders used to identify WL voids within WL convergence or peak fields. The void finders discussed in this work are modified versions of popular void finders that are typically applied to the galaxy distribution. We have shown how each void finder can be modified such that it can be applied to WL maps and have discussed the impact of varying each free parameter associated with the void finders (see Section \ref{sec:void finders}). The WL void finders have been split broadly into two classes: i) those that can identify voids directly in the WL convergence maps, and ii) those that require WL peaks as tracers in order to define the voids. We have found that both void classes offer useful information.

We investigate the WL void abundances, convergence profiles and tangential shear profiles for all void finders (where applicable) in Section \ref{sec:void statistics}. The average void convergence profile consists of an underdense region (i.e. $\kappa<0$) for $r\lesssim R_{\rm{v}}$ (with $R_{\rm{v}}$ the void radius), an overdensity at $r\sim R_{\rm{v}}$ (not present for troughs), followed by a slow convergence to the background expectation of $\kappa=0$ at large radial distances. This translates into a negative tangential shear profile for voids, with the amplitude of $\gamma_{\rm{t}}$ being maximal at $r\simeq R_{\rm{v}}$. We found that WL minima and SVF $\kappa$ produce the deepest (most underdense) convergence profiles at $r = 0$, and the $\gamma_{\rm{t}}$ profiles with the largest amplitudes are produced by tunnels (without GSN) and WL minima (with GSN). 

To differentiate the various void finders, we have studied, for each void type, the impact of GSN and the SNR with which their tangential shear profiles can be measured in an \LSST{} like survey. In general, voids identified directly in the convergence field have the highest $\gamma_{\rm{t}}$ SNR but are also most severely affected by GSN. The void finders based on the peak distribution have moderate SNR and are less affected by GSN. Troughs with large sizes are least impacted by GSN but are also the ones with the lowest $\gamma_{\rm{t}}$ SNR. Increasing the smoothing length or the peak threshold used to identify voids, while it lowers the impact of GSN,  also decreases the SNR with which the void tangential shear profile can be measured. The tunnel algorithm provides a good compromise between mitigating the impact from GSN and producing objects with a large $\gamma_{\rm{t}}$ SNR.

In a future work we will use WL voids to provide cosmological parameter constraints and investigate how WL void statistics can be used in a manner that is complementary to constraints from other probes such as WL peaks and the convergence power spectrum. This will be especially interesting in the context of the $\Omega_{\rm{m}} -\sigma_8$ degeneracy. Both galaxy voids and WL peaks have been shown to be able to help break this parameter degeneracy \citep{Nadathur2019,Dietrich2010,Davies2019}, and WL voids may offer another promising avenue to do so.

For parameter constraints, tunnels may prove useful, since we have found it to be the best WL void finder working in the WL peak distribution, in terms of both large SNR value and small impact from GSN, followed closely by SVF peaks. The high SNR values from the WL minima and WVF tangential shear profiles make these WL void definitions viable candidates for parameter constraints as well. It is possible that void finders applied directly to the convergence field may be complementary to those that use WL peaks, since they are sensitive to different aspects of the WL convergence maps when identifying voids.

Additionally, some of the void finders have high SNR values for all smoothing scales studied here. This makes combining different smoothing scales a possible and potentially useful approach when applied to cosmological parameter constraints, since it has been shown that constraints from WL peaks are improved when multiple smoothing scales are used \citep{J.Liu2015}. Finally, in this work we discuss the merit of a given WL void in terms of their tangential shear profiles, however other WL void statistics such as the void abundance and void correlation functions may also provide useful cosmological information. 

When considering the impact of baryons on the WL void statistics, sufficiently large smoothing scales must be used in order to get agreement between hydro simulations and dark matter only simulations, as is the case with other WL statistics \citep{Weiss2019}. \cite{Paillas2017} have shown that voids in the LSS are less impacted by baryons, and \cite{Coulton2019} have shown that WL minima are more robust to baryons than WL peaks. Therefore, given that \cite{Chang2018} have also shown that the deepest WL minima correspond to large supervoids, confirming that the underdense regions of the WL convergence maps are due to underdensities along the line of sight, it is reasonable to expect that the WL voids identified directly in the convergence maps may be more resilient to baryonic physics. However, the void finders which use WL peaks as tracers will be more affected since WL peaks are more sensitive to baryons \citep{Osato2015,Weiss2019,Coulton2019}, and changes to the WL peak distribution could impact the resulting void catalogues. More detailed studies, potentially with the aid of cosmological hydrodynamic simulations, are needed to better understand these issues.

\section*{Acknowledgements}

CTD is funded by a UK Science and Technology Facilities Council (STFC) PhD studentship through grant ST/R504725/1. EP is supported by CONICYT-PCHA/Doctorado Nacional (2017-21170093) and also acknowledges support from CONICYT project Basal AFB-170002. MC is supported by the EU Horizon 2020 research and innovation programme under a Marie Sk{\l}odowska-Curie grant agreement 794474 (DancingGalaxies). BL is supported by an ERC Starting Grant, ERC-StG-PUNCA-716532, and additionally supported by the STFC Consolidated Grants [ST/P000541/1, ST/T000244/1].

This work used the DiRAC@Durham facility managed by the Institute for Computational Cosmology on behalf of the STFC DiRAC HPC Facility (www.dirac.ac.uk). The equipment was funded by BEIS capital funding via STFC capital grants ST/K00042X/1, ST/P002293/1, ST/R002371/1 and ST/S002502/1, Durham University and STFC operations grant ST/R000832/1. DiRAC is part of the National e-Infrastructure.

\section{Data Availability}

The data used in this work is publicly available at \href{http://cosmo.phys.hirosaki-u.ac.jp/takahasi/allsky_raytracing/}{http://cosmo.phys.hirosaki-u.ac.jp/takahasi/allsky\_raytracing/}




\bibliographystyle{mnras}
\bibliography{mybib} 



\appendix
\section{Correlation matrices}
\label{app:correlation}

\begin{figure*}
    \centering
    \includegraphics[width=2\columnwidth]{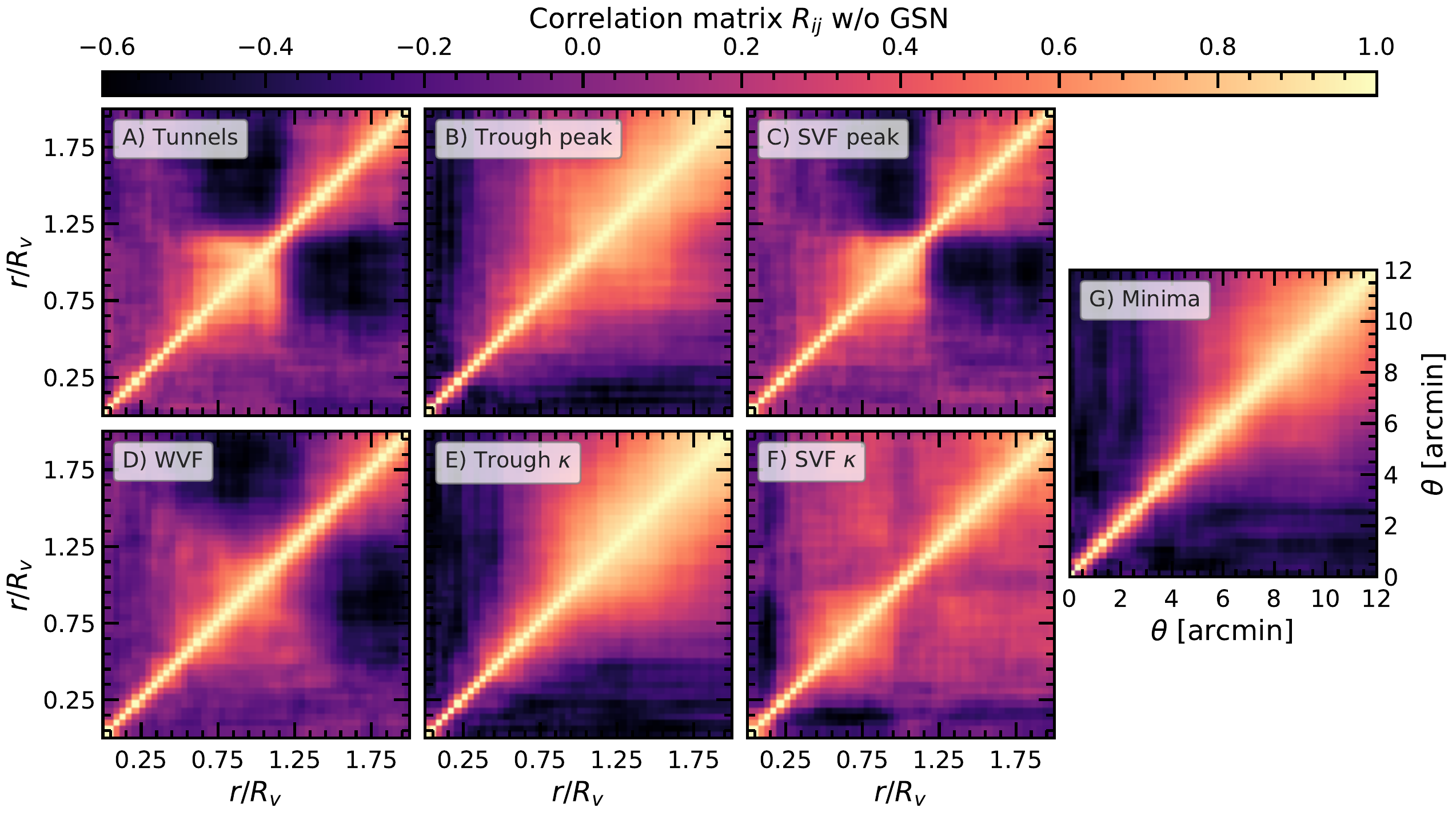}
    \caption{The tangential shear correlation matrices for all void finders discussed in this work, calculated from maps with no GSN and smoothed with $\theta_{\rm{s}} = 2.5$ arcmin. For WL void finders applied to the WL peak distribution, results are present for peak catalogues with $\nu > 2$.}
    \label{fig:correlation noGSN}

    \includegraphics[width=2\columnwidth]{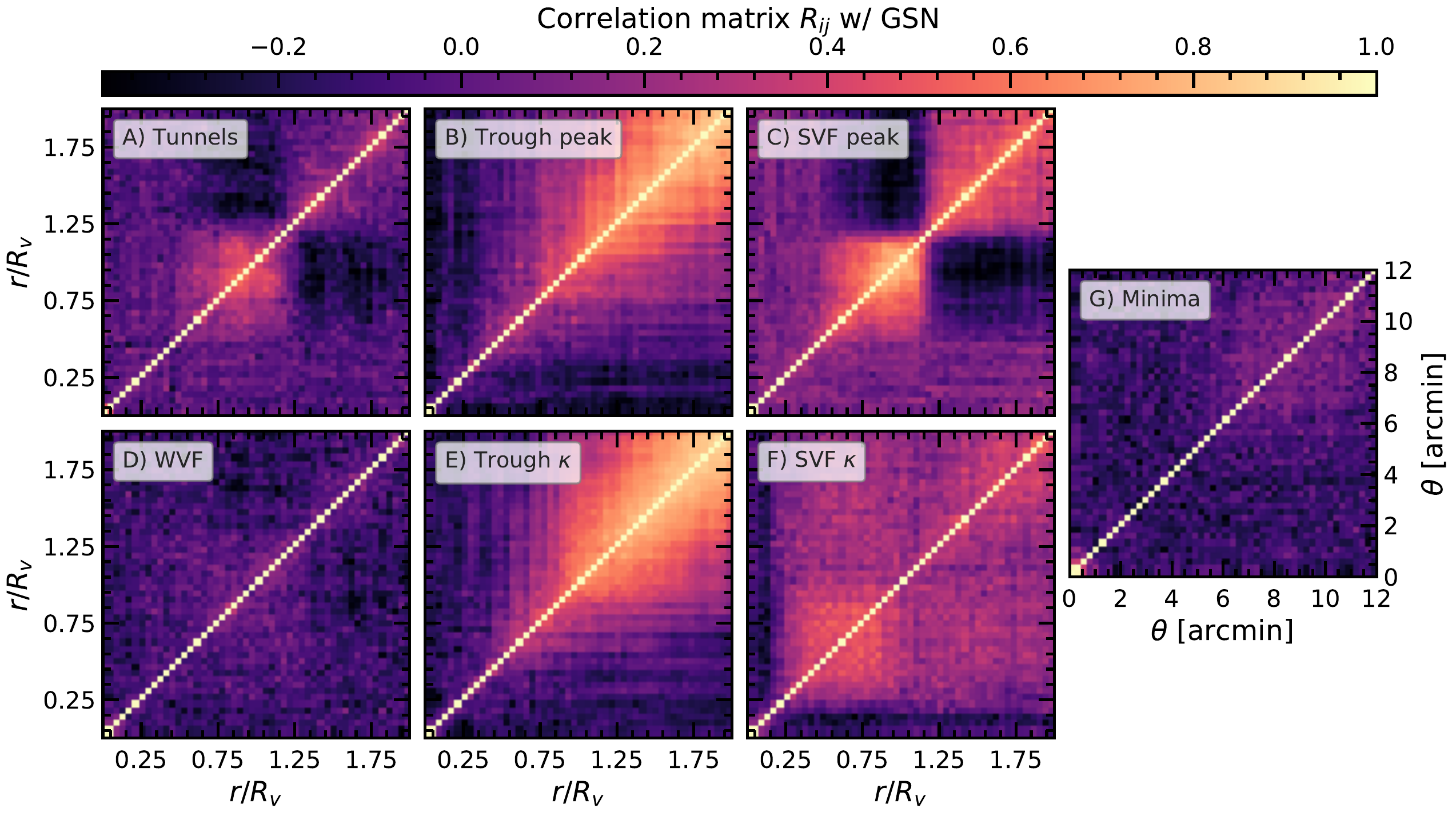}
    \caption{The same as Fig.~\ref{fig:correlation noGSN} but for convergence maps that include GSN.}
    \label{fig:correlation GSN}
\end{figure*}

In this Appendix we present the tangential shear correlation matrices for the void finders we have studied. For simplicity we present all correlation matrices for a smoothing scale of $\theta_{\rm{s}}=2.5$ arcmin and for peak catalogues with $\nu>2$ where applicable. Fig.~\ref{fig:correlation noGSN} shows the tangential shear correlation matrices for WL voids identified in WL maps without GSN, and Fig.~\ref{fig:correlation GSN} is the same but for WL maps with GSN included. The correlation matrix $R_{ij}$ is related to the covariance matrix, ${\rm cov}_{ij}$ (which is used to calculate SNR values in Eq.~\eqref{eq:SNR}), through the equation,
\begin{equation}
    R_{ij} = \frac{{\rm Cov}_{ij}}{\sigma_i\sigma_j} \, ,
    \label{eq:correlation matrix}
\end{equation}
where $i$ and $j$ are radial bin indices, $R$ is the correlation matrix, $\rm{cov}$ is the covariance matrix and $\sigma_i$ is the standard deviation in bin i, where the variance, $\sigma^2$, is given by the diagonal elements of the covariance matrix. The covariance matrix is calculated as
\begin{equation}
    {\rm Cov}_{ij} = \frac{1}{N-1} \sum_{k=1}^{N} [\gamma_{\rm{t}}(i) - \bar{\gamma}_{\rm{t}}(i)][\gamma_{\rm{t}}(j) - \bar{\gamma}_{\rm{t}}(j)] \, ,
\end{equation}
where $N = 192$ is the number of WL maps, $\gamma_{\rm{t}}$ the tangential shear, and an over-bar denotes the mean from $N$ maps. 

Fig.~\ref{fig:correlation noGSN} shows the $\gamma_{\rm{t}}$ correlation matrices for maps without GSN. The seven panels correspond to the seven WL void finders studied in this work, where dark colours indicate an anti-correlation between bins and bright colours indicate a correlation between bins (as indicated by the colour bar). In all cases, the region around the diagonal is close to unity, illustrating that neighbouring bins are highly correlated. Of all the void finding algorithms, the ones with the most correlated bins appear to be the two trough finders. This is due to the large degree of overlap between neighbouring troughs as seen in Fig.~\ref{fig:visualisation}, and it is this correlation between far apart bins that produces a lower SNR for the trough algorithms relative to the other void finders in the right panel of Fig.~\ref{fig:diff_and_SNR}. Similarly, Fig.~\ref{fig:visualisation} also shows that the SVF $\kappa$ voids tend to clump together and overlap with each other, which explains why there is also a significant correlation between different radial bins. The same happens, though to a lesser extent, to WL minima, because there is a large number of them and so the large radius bins (of which the radii become a substantial fraction of the inter-minimum separation) start to overlap between neighbouring minima.

Fig.~\ref{fig:correlation GSN} is the same as Fig.~\ref{fig:correlation noGSN}, except that here we study void populations identified in WL maps that include GSN. The correlation matrices are significantly more diagonal when GSN is included, which shows that GSN reduces the correlation between all bins; this is partly responsible for the increase in SNR when GSN is included as shown in the right panel of Fig.~\ref{fig:diff_and_SNR}. Since GSN does not reduce the amplitude of the tangential shear profiles, but does reduce the covariance between different bins, the $\gamma_{\rm{t}}$ and $\rm{Cov}^{-1}$ terms in Eq.~\eqref{eq:SNR} increase, yielding a larger SNR. Despite the reduction in correlation between bins from GSN, the troughs algorithms, and to a lesser extent SVF $\kappa$, still have show a considerable correlation between bins with $r\gtrsim R_{\rm{v}}$, which again is due to many troughs overlapping in maps with GSN. Interestingly, adding GSN seems to reduce the correlation between different bins more efficiently for tunnels than for SVF peak voids. Finally we have checked and verified that the covariance matrices presented here agree with covariance matrices calculated from a bootstrapped version of our data set.

\section{WL voids in GSN only maps}
\label{app:WL voids in GSN maps}

\begin{figure*}
    \centering
    \includegraphics[width=2\columnwidth]{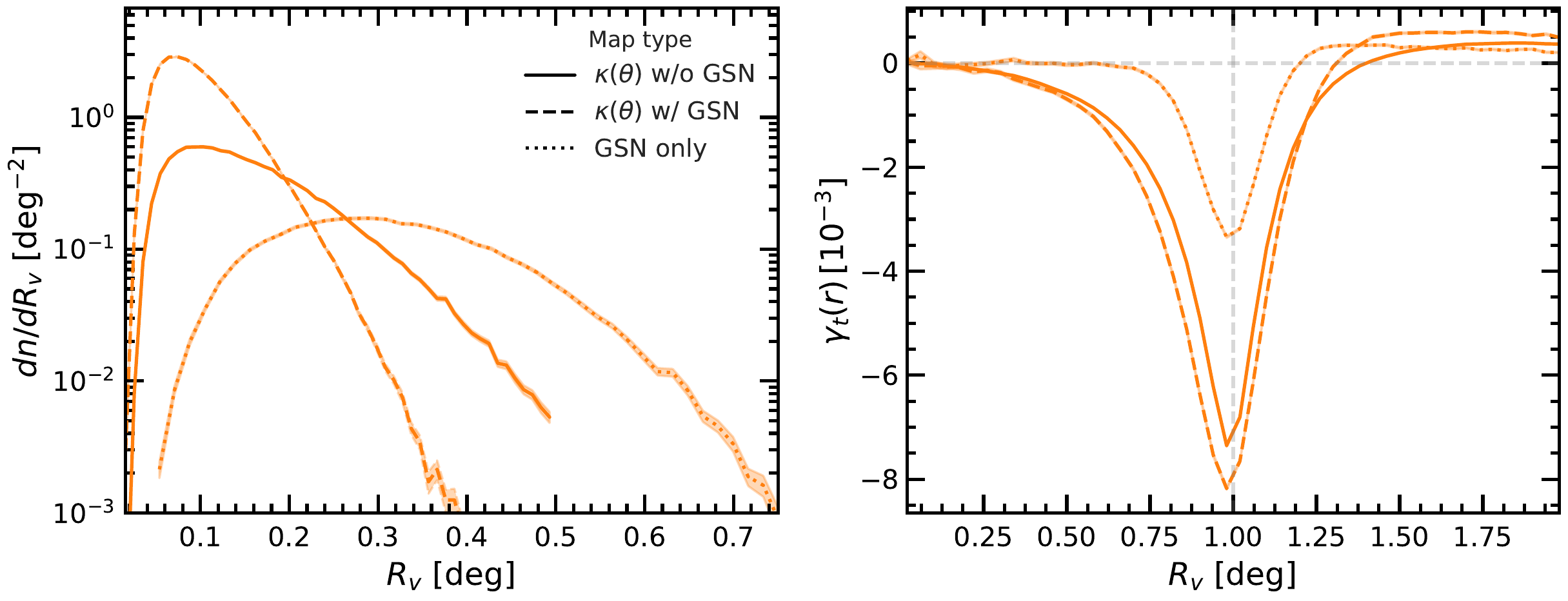}
    \caption{Tunnels identified in three different convergence maps: physical convergence without GSN (solid), physical convergence with GSN added (dashed), and GSN-only (dotted). The tunnels are identified in peak catalogues with heights $\nu > 2$, and using a smoothing scale $\theta_{\rm{s}} = 2.5$ arcmin. The shaded regions around the lines indicate the one sigma standard error bars.}
    \label{fig:noise_only}
\end{figure*}

Typically, 3D voids in the LSS are identified in galaxy distributions, where galaxies are used as tracers for a given void finder. The void lensing signal is then extracted from lensing measurements that are separate from the galaxy position measurements. This means that the observational noise and systematics associated with the galaxy positions are (mostly) independent of the noise and systematics in the lensing measurements.

In the case of WL voids, the same measurement (the WL convergence map) is used to identify voids and to measure their lensing profiles. This means that the void identification process and void lensing profiles will be closely connected, and impacted by noise in similar ways. The connection between WL void identification and the corresponding lensing profiles can be further strengthened by the fact that each void finder yields distinct lensing profile shapes that are determined by the definition employed to identify the voids, as shown and discussed in Section \ref{sec:void statistics}. Taking tunnels as an example: because by definition each tunnel contains no WL peaks and has at least three peaks on its boundary, we should expect the convergence profile to have a peak at the tunnel radius, being negative inside and approaching the background value far away (i.e., the same qualitative behaviour as seen in the physical signal in Fig.~\ref{fig:tunnels statistics}), even if the peaks are identified from a pure noise map. In other words, the WL void lensing profiles could simply be a consequence of the way 2D voids are identified from {\it any} WL convergence or peak distribution, rather than a physical effect.

Given that observed WL convergence maps are significantly contaminated by GSN, this means that voids identified in WL maps could potentially be due to noise, or they could be indistinguishable from spurious voids that result from noise. It is therefore important to understand how to distinguish between voids that are produced by physical signals in the WL maps and spurious voids that are the result of noise.
This is the primary reason why in this paper we have tried to smooth the WL maps using filters as large as $5$ arcmin, in order to suppress the impact of GSN on the measured peak and void statistics, so that the results from the no-GSN and GSN-added maps agree with each other. For completeness, in this appendix we give a slightly more detailed comparison, where we show how WL void statistics behave when these void finders are applied directly to a noise map, which is a mock WL map which contains no physical signal whatsoever.

In order to generate a GSN-only WL map, we follow the same GSN prescription used throughout this work. We first define a grid of pixels which matches the same angular size and resolution of the WL maps used in the rest of this work, and set the value of each pixel to zero. For each pixel we then add randomly drawn values from the Gaussian distribution described in Section \ref{sec:GSN prescription}, Eq.~\eqref{eq: GSN gaussian}.

Fig.~\ref{fig:noise_only} shows tunnels identified in three WL maps: without GSN (solid), with GSN (dashed) and GSN-only (dotted). The results shown correspond to a smoothing scale, $\theta_{\rm{s}} = 2.5$ arcmin, and are obtained using WL peaks with heights, $\nu > 2$. The left panel shows the abundances of the tunnels in the three map types. The GSN-only maps produce fewer tunnels, which are typically larger than the tunnels in the physical maps. In particular, the GSN-only maps produces fewer small voids and more large voids, when compared to the other two map types. This results from WL peaks clustering in the maps that contain a physical signal, and thus many of peaks are close together and produce smaller tunnels. Whereas the GSN-only maps have fewer peaks that by definition do not cluster, which results in larger voids. 

The right panel shows the tangential shear profiles for tunnels identified in the three map types. As shown by the dotted line, the tangential shear profiles for the GSN-only maps remain flat at $\gamma_{\rm{t}} = 0$ for most of the void interior, where departure from zero only occurs near the void boundary at $r\sim0.75 R_{\rm{v}}$. This is due to the fact that the void interiors in the GSN-only maps are on average not underdense, which in turn is because of the random nature of the pure GSN map and the lack of gravity to physically evacuate matter from the void. Furthermore, the amplitude of $\gamma_{\rm{t}}$ at $r \simeq R_{\rm{v}}$ is significantly lower than for the maps that 
contain the physical signal. This is due to noise-only tunnels having less overdense boundaries than their physical counterparts. This can be understood as follows. For the noise-only maps, the three peaks which determine the tunnel boundary are overdense, but, since different points in noise-only maps are uncorrelated, the remaining pixels along the boundary can take any values and thus they would have a mean convergence of $0$. In contrast, the correlations present in the physical maps mean that the pixels found at the boundary of physical tunnels are on average overdense since they are close to the overdense peaks used to define the tunnel.

We also find that the $\gamma_{\rm{t}}$ profile for WL tunnels identified in a pure noise map is much more sensitive to the smoothing scale $\theta_{\rm{s}}$ used to smooth the convergence map. Although not shown here for the sake of clarity, we have checked the cases $\theta_{\rm{s}}=1$ and $5$ arcmin respectively. In the former case, the peak of the tangential shear profile from the pure noise map is as deep as that from the physical WL map, whereas in the latter case, the peak of the tangential shear profile from the pure noise map is further suppressed and becomes very weak. The same is found for WL peak catalogues with other $\nu$ thresholds.

It is evident from these tests that the statistics used to describe WL voids in this work give distinct results for the GSN-only maps, relative to the physical WL maps. This shows that WL voids are sensitive to the physical information present in WL maps, even when GSN is included.


\bsp	
\label{lastpage}
\end{document}